{}
{}

\documentclass[12pt,a4paper]{article}

\newif\ifpublic\publicfalse
\newif\ifniklas\niklastrue

\usepackage{ifpdf}
\ifniklas\ifpdf\publictrue\else\publicfalse
\PassOptionsToPackage{hypertex}{hyperref}
\PassOptionsToPackage{draft}{graphicx}
\fi\fi
\usepackage{amsmath,amssymb}
\usepackage[bookmarks=true,hyperfigures=true]{hyperref}
\usepackage{graphicx}
\usepackage[nosort]{cite}
\usepackage[bulletsep]{collref}
\usepackage{graphicx}
\usepackage[usenames,dvipsnames]{color}

\ifniklas

\def\showkeysrefformat#1{{\normalfont\tiny\ttfamily#1}}
\makeatletter
\def\SK@@ref#1>#2\SK@{%
 {\@inlabelfalse\leavevmode\vbox to\z@{%
 \vss\SK@refcolor\rlap{\vrule\raise .75em%
  \hbox{\showkeysrefformat{#2}}}}}}
\makeatother
\fi

\usepackage[a4paper,text={161mm,247mm},centering]{geometry} 

\allowdisplaybreaks[3]

\numberwithin{equation}{section}

\usepackage[font=small,labelfont=bf,width=0.85\textwidth]{caption}

\expandafter\def\expandafter\bfseries\expandafter{\bfseries\ifmmode\else\boldmath\fi}
\expandafter\def\expandafter\mdseries\expandafter{\mdseries\ifmmode\else\unboldmath\fi}
\expandafter\def\expandafter\normalfont\expandafter{\normalfont\ifmmode\else\unboldmath\fi}

\RequirePackage{verbatim}

\makeatletter
\newwrite\bibinl@out
\newenvironment{bibtex}[1][\jobname]{%
  \immediate\openout\bibinl@out #1.bib
  \immediate\write\bibinl@out{\@percentchar generated from `\jobname' starting line \the\inputlineno^^J}%
  \def\verbatim@processline{\immediate\write\bibinl@out{\the\verbatim@line}}%
  \@bsphack\let\do\@makeother\dospecials\catcode`\^^M\active\verbatim@start
}%
{\immediate\closeout\bibinl@out\@esphack}
\makeatother

\RequirePackage{verbatim}

\makeatletter
\newwrite\mpi@out
\def\mpi@write#1{\immediate\write\mpi@out{#1}}
\immediate\openout\mpi@out\jobname.mp
\mpi@write{\@percentchar generated from `\jobname'}%
\AtEndDocument{\immediate\closeout\mpi@out}

\newcommand{\mpi@putlineno}{%
  \mpi@write{\@percentchar---------------------------------------}%
  \mpi@write{\@percentchar l.\the\inputlineno}%
}

\newcommand{\mpi@verbatim}{
  \@bsphack
  \let\do\@makeother\dospecials
  \catcode`\^^M\active
  \def\verbatim@processline{\mpi@write{\the\verbatim@line}}%
  \verbatim@start
}

{\mpi@write{}\@esphack}

{\mpi@write{endfig;}\@esphack}

\newcommand{\includegraphicsex}[2][]{%
  \xdef\mpi@tmp{#2}%
  \IfFileExists{\mpi@tmp}%
    {\includegraphics[#1]{\mpi@tmp}}%
    {\textbf{??}\typeout{file \mpi@tmp{} missing}}%
}

\makeatother

\ifx\genfrac\sdflkaj\else\fi
\newcommand{\sfrac}[2]{{\textstyle\frac{#1}{#2}}}
\newcommand{\half}{\sfrac{1}{2}}

\newcommand{\hopf}[1]{\mathrm{#1}}
\newcommand{\yang}{\hopf{Y}}

\newcommand{\alg}[1]{\mathfrak{#1}}
\newcommand{\grp}[1]{\mathrm{#1}}

\newcommand{\sprods}[2]{\langle#1#2\rangle}

\newcommand{\cprods}[2]{[#1#2]}

\newcommand{\superN}{\mathcal{N}}

\newcommand{\nln}{\nonumber\\}

\def\[{\begin{equation}}
\def\]{\end{equation}}

\providecommand{\href}[2]{#2}

\makeatletter
\def\mr@ignsp#1 {\ifx\:#1\@empty\else #1\expandafter\mr@ignsp\fi}%
\newcommand{\multiref}[1]{\begingroup
\xdef\mr@no@sparg{\expandafter\mr@ignsp#1 \: }%
\def\mr@comma{}%
\@for\mr@refs:=\mr@no@sparg\do{\mr@comma\def\mr@comma{,}\ref{\mr@refs}}%
\endgroup}
\renewcommand{\eqref}[1]{(\multiref{#1})}
\makeatother

\makeatletter
\newcommand{\namedref}[2]{\hyperref[#2]{#1~\ref*{#2}}}
\newcommand{\secref}{\@ifstar{\namedref{Section}}{\namedref{sec.}}}
\newcommand{\subsecref}{\@ifstar{\namedref{Subsection}}{\namedref{subsec.}}}
\newcommand{\appref}{\@ifstar{\namedref{Appendix}}{\namedref{app.}}}
\newcommand{\tabref}{\@ifstar{\namedref{Table}}{\namedref{tab.}}}
\newcommand{\figref}{\@ifstar{\namedref{Figure}}{\namedref{fig.}}}
\makeatother

\providecommand{\hypersetup}[1]{}

\hypersetup{plainpages=false}
\hypersetup{pdfpagemode=UseNone}
\hypersetup{bookmarksnumbered=true}
\hypersetup{pdfstartview=FitH}
\hypersetup{colorlinks=false}
\hypersetup{citebordercolor={.5 1 .5}}
\hypersetup{urlbordercolor={.5 1 1}}
\hypersetup{linkbordercolor={1 .7 .7}}

\makeatletter
\let\@keywords\@empty
\let\@subject\@empty
\providecommand{\keywords}[1]{\gdef\@keywords{#1}}
\providecommand{\subject}[1]{\gdef\@subject{#1}}
\def\thetitle{\@title}
\def\theauthor{\@author}
\def\thesubject{\@subject}
\def\thedate{\@date}
\def\thekeywords{\@keywords}
\makeatother
\AtBeginDocument{
\hypersetup{pdftitle={\thetitle}}%
\hypersetup{pdfauthor={\theauthor}}%
\hypersetup{pdfsubject={\thesubject}}%
\hypersetup{pdfkeywords={\thekeywords}}%
}

\makeatletter
\newsavebox{\apb@box}\newlength{\apb@width}
\newcommand{\autoparbox}[2][c]{\sbox{\apb@box}{#2}%
 \settowidth{\apb@width}{\usebox{\apb@box}}%
 \parbox[#1]{\apb@width}{\usebox{\apb@box}}}

\newcommand{\includegraphicsboxex}[2][]{\autoparbox{\includegraphicsex[#1]{#2}}}
\makeatother

\newif\ifmrnote 
\mrnotetrue

\usepackage[active]{srcltx}
\usepackage{color}

\newcommand{\rop}{\mathrm{R}}
\newcommand{\lop}{\mathrm{L}}
\newcommand{\dl}[1]{\delta_{#1}}
\newcommand{\dlb}[1]{\delta_{\text{-}#1}}

\newif\ifjbnote 
\jbnotetrue

\newenvironment{permutation}{\Big(\begin{smallmatrix}}{\end{smallmatrix}\Big)}

\title{Deformed one-loop amplitudes in\\$\superN=4$ super-Yang--Mills theory}
\author{Johannes Broedel, Marius de Leeuw and Matteo Rosso}

\begin{document}

\iftrue

\pdfbookmark[1]{Title Page}{title}
\thispagestyle{empty}


\vspace*{2cm}
\begin{center}%
\begingroup\Large\bfseries\thetitle\par\endgroup
\vspace{1cm}

\begingroup\scshape\theauthor\par\endgroup
\vspace{5mm}%

\begingroup\itshape
Institut f\"ur Theoretische Physik,\\
Eidgen\"ossische Technische Hochschule Z\"urich\\
Wolfgang-Pauli-Strasse 27, 8093 Z\"urich, Switzerland
\par\endgroup
\vspace{5mm}

\begingroup\ttfamily
\verb+{+jbroedel,deleeuwm,mrosso\verb+}+@itp.phys.ethz.ch
\par\endgroup

\vfill

\textbf{Abstract}\vspace{5mm}

\begin{minipage}{12.7cm}
  We investigate Yangian-invariant deformations of one-loop amplitudes in
  $\superN=4$ super-Yang--Mills theory employing an algebraic representation of
  amplitudes. In this language, we reproduce the deformed massless box integral
  describing the deformed four-point one-loop amplitude and compare different
  realizations of said amplitude.
\end{minipage}

\vspace*{4cm}

\end{center}

\newpage

\fi

\tableofcontents

\section{Introduction and outline}
\label{sec:intro}

In a series of recent papers, deformations of Yangian invariants in the context
of $\superN=4$ super-Yang--Mills (sYM) theory have been investigated
\cite{Ferro:2012xw,Ferro:2013dga,Beisert:2014qba,Kanning:2014maa,Broedel:2014pia}.
As opposed to the undeformed situation, a deformed Yangian invariant allows for
nonzero expectation values of the central charge operator for each external leg
-- which amounts to shifting the helicities of the external particles.
Including the hypercharge as well, the underlying symmetry algebra is extended
from the Yangian $\yang[\alg{psu}(2,2|4)]$ to $\yang[\alg{u}(2,2|4)]$.  

Yangian invariance, however, constrains the allowed deformations by linking the
central charges of the external legs to the evaluation parameters of an
evaluation representation of the Yangian algebra, thereby encoding a
permutation as discussed in refs.~\cite{Kanning:2014maa,Broedel:2014pia}. At
tree level, the permutation labels a Yangian invariant unambiguously and can be
translated into on-shell graphs \cite{ArkaniHamed:2012nw} and $\rop$-operators
\cite{Chicherin:2013ora}.

The relation of deformed Yangian invariants to scattering amplitudes in
$\superN=4$ sYM theory has been discussed in
refs.~\cite{Beisert:2014qba,Kanning:2014maa,Broedel:2014pia}. All tree-level
amplitudes in the maximally-helicity-violating (MHV) sector are represented by
a single Yangian invariant, and can thus be deformed. For tree-level amplitudes
of higher MHV degree this is not possible any more, as those are composed from
several Yangian invariants.  The obstruction here is physicality, which demands
compatibility between all Yangian invariants contributing to the amplitude: all
external legs should have the \textit{same} data associated to them\footnote{The
six-point NMHV amplitude is an exception, as will be explained in
\secref{sec:ampl} below.}.   

Deformations of the four-point one-loop amplitude have been considered in
refs.~\cite{Ferro:2012xw,Ferro:2013dga,Beisert:2014qba}. In parallel to the
tree-level situation, the integrand of the amplitude is a Yangian invariant
only for certain deformations. For deformations not leading to a Yangian
invariant, one can perform the integration and obtain a result which is
reminiscent of the usual expression for the undeformed four-point one-loop
amplitude \cite{Ferro:2012xw,Ferro:2013dga}. This fact suggests to use the
deformation as a regulator similar to analytic regularization.
Keeping Yangian invariance in the integrand, that is, choosing a
Yangian-invariant deformation, renders the integration difficult and leads to a
vanishing result except for very special deformations \cite{Beisert:2014qba}.

In this note, we investigate how to take deformations to the loop level in a
natural way by employing the $\rop$-operator formalism described in
ref.~\cite{Chicherin:2013ora}. We will discuss general features of bubbles in
on-shell diagrams in the language of $\rop$-operators, which allows to treat
the momentum-space properties as well as the Yangian properties simultaneously.
From this minimal example, which already exhibits all features of loop
amplitudes, we will proceed to deformations of the four-point one-loop
integrand of ref.~\cite{ArkaniHamed:2012nw} in the language of
$\rop$-operators. 

In particular we compare this deformed amplitude with the four-point one-loop
amplitude discussed in refs.~\cite{Ferro:2012xw,Ferro:2013dga,Beisert:2014qba}.
While both integrals lead to the same deformed momentum-space integral, the
$\rop$-operator language reveals an astonishing fact: they constitute different
eigenstates of the monodromy matrix. Thus, either the eigenvalues have to
agree, which will be shown to lead to the trivial deformation or the eigenstate
has to vanish or diverge. The vanishing result is in agreement with the
conclusion drawn in ref.~\cite{Beisert:2014qba}.

So the unregulated\footnote{We will use the term \textit{loop amplitude} below
in order to label the Yangian invariant, that is, without performing the
integrations. Nevertheless, usually one would refer to the regulated result in
a particular regularization scheme as the loop amplitude.} loop amplitude
\textit{is} a Yangian invariant: it is either zero or infinity. Yangian
invariance is only broken by regulating the integral: the infrared divergences
arising during the regularization process introduce a scale and thus break
conformal invariance. 

After the discussion of the four-point one-loop case, we turn our attention to
the five-point one-loop amplitude. Based on the Britto--Cachazo--Feng--Witten
(BCFW) construction \cite{Britto:2004ap,Britto:2005fq} of loop integrands
introduced in ref.~\cite{ArkaniHamed:2010kv}, we consider deformations of the
three contributing BCFW-channels in order to arrive at what seems to be a
general statement: for loop integrands constructed from several Yangian
invariants, there is no consistent deformation if one requires physicality.
This argument is analogous to the one used in ref.~\cite{Broedel:2014pia} for
tree-level amplitudes in the NMHV sector. For higher-loop amplitudes the
situation does not improve.

After a brief review of the necessary techniques for the investigation
of deformed scattering amplitudes in \secref{sec:ampl} we collect previous
results on four-point one-loop amplitudes in \secref{sec:oneloopreview}.
\secref*{sec:bubbles} is devoted to the discussion of the integrals
occurring in loop constructions on the simple example of a bubble-shaped
on-shell graph. In \secref{sec:oneloopcalc} we finally use the $\rop$-operator
formalism in order to build the four-point one-loop amplitude. We investigate the
resulting integral and its branch cut structure in a deformed scenario in order
to show that the deformation renders the integral trivial.  In
\secref{sec:onshell}, we construct the five-point one-loop amplitude following
ref.~\cite{ArkaniHamed:2010kv} and comment on possible deformations. 


\section{Amplitudes in $\superN=4$ sYM theory, on-shell diagrams and $\rop$-operators}
\label{sec:ampl}

In this section we are going to review two descriptions of Yangian invariants
relevant in the context of scattering amplitudes in planar $\superN=4$ sYM
theory: on-shell diagrams and the $\rop$-operator formalism. The relations
between these two formulations and permutations, which can be used to uniquely
label tree-level Yangian invariants, have been thoroughly investigated
in refs.~\cite{Kanning:2014maa,Broedel:2014pia}. Here we will be
rather brief and refer the reader to these references for further details.

Both descriptions of Yangian invariants, $\rop$-operators as well as on-shell
diagrams, rely on the on-shell superspace \cite{Nair:1988bq} with variables
$(\lambda^\alpha,\tilde{\lambda}_{\dot{\alpha}},\tilde{\eta}^A)$, where Greek
and upper case Latin indices label the fundamental representations of
$\grp{SL}(2)$ and $\grp{SU}(4)$ respectively.

Yangian invariants with $n$ external legs are functions defined on the $n$-fold
tensor product of on-shell superspace and are annihilated by all generators of
the Yangian algebra (see ref.~\cite{Beisert:2014qba} for a short review of
Yangian algebras in this context). Extending the algebra $\alg{psu}(2,2|4)$ --
which is the algebra underlying planar $\superN=4$ sYM theory -- with the
central charge operator $\alg{C}$ and the hypercharge $\alg{B}$
yields\footnote{At the level of the Yangian, the hypercharge $\alg{B}$ is a
symmetry \cite{Beisert:2011pn}. } the algebra $\alg{u}(2,2|4)$. Up to reality
conditions, the algebra $\alg{u}(2,2|4)$ is equal to $\alg{gl}(4|4)$, which we
will be concerned with below. Choosing furthermore an evaluation representation
for $\yang[\alg{u}(2,2|4)]$ with evaluation parameters $u_i$ leads to a set
$(\lambda^\alpha_i,\tilde{\lambda}_{i\dot{\alpha}},\tilde{\eta}^A_i, c_i, u_i)$
of external data for each leg, where $c_i$ is the eigenvalue of the central
charge operator $\alg{C_i}$. The total number of $\eta$'s in a $n$-point
Yangian invariant determines the variable $k$ labeling the MHV-sector via
$k=\sfrac{\#\eta\text{'s}}{4}$ \cite{Nair:1988bq}.

\begin{figure}
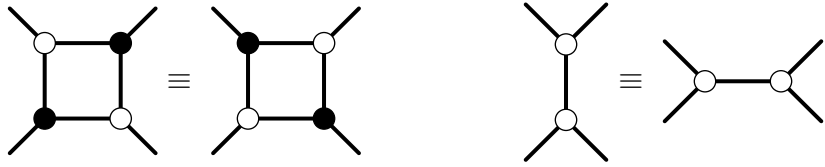

  \centering
  \includegraphicsboxex{Fig_sq1.mps}$\;\equiv\;$
  \includegraphicsboxex{Fig_sq2.mps}\hspace{2cm}
  \includegraphicsboxex{Fig_merg1.mps}$\;\equiv\;$
  \includegraphicsboxex{Fig_merg2.mps}
  \caption{Square move and merger.}
  \label{fig:moves}
\end{figure}

On-shell diagrams representing invariants for $\yang[\alg{u}(2,2|4)]$ are
composed by gluing two types of deformed three-point vertices
\cite{Ferro:2012xw,Ferro:2013dga}
\begin{align}
  \label{eq:3val_bl}
  \mathcal{A}_\bullet = \frac{\delta^4 (P)\delta^8
    (Q)}{\sprods{1}{2}^{1+c_3} \sprods{2}{3}^{1+c_1}
    \sprods{3}{1}^{1+c_2}},\qquad
  \mathcal{A}_\circ = \frac{\delta^4 (P)
    \delta^4(\tilde{Q})}{\cprods{1}{2}^{1-c_3}
    \cprods{2}{3}^{1-c_1} \cprods{3}{1}^{1-c_2}}\,,
\end{align}
where $P:=\sum_{i=1}^3 \lambda^\alpha_i \tilde{\lambda}^{\dot{\alpha}}_i$
denotes the total four-momentum, whereas $Q:=\sum_i \lambda_i^\alpha
\tilde{\eta}_i^A$ and $\tilde{Q}:=\cprods{1}{2} \tilde{\eta}_3^A
+\cprods{2}{3}\tilde{\eta}_1^A+\cprods{3}{1} \tilde{\eta}_2^A$. Each of the
building blocks $\mathcal{A}_\bullet$ and $\mathcal{A}_\circ$ is a Yangian
invariant if and only if the following equations are satisfied:
\begin{align}
  \label{eq:z_to_u_b}
  \mathcal{A}_\bullet:& \qquad c_1 = u_1-u_3,\quad c_2= u_2 -
  u_1, \quad c_3 = u_3 - u_2\,; \\
  \label{eq:z_to_u_w}
  \mathcal{A}_\circ:&\qquad c_1 = u_1-u_2,\quad c_2= u_2 - u_3, \quad
  c_3 = u_3 - u_1\,,
\end{align}
which implies
$\sum_i\alg{C}_i\mathcal{A}_\bullet=\sum_i\alg{C}_i\mathcal{A}_\circ=0$.
Combining several of those building blocks in a Yangian-invariant way by
gluing leg $V$ to leg $W$ requires (see ref.~\cite{Beisert:2014qba} for a
derivation\footnote{Notice, however, that we are using the conventions 
of ref.~\cite{Broedel:2014pia} here.})
\[
  \label{eq:gluing_cond}
  c_V=-c_W \quad\text{as well as}\quad u_V-\half c_V = u_W-\half c_W.
\]
An on-shell graph represents a Yangian invariant if and only if the system of
equations composed from all vertex conditions (cf. eqns.~\eqref{eq:z_to_u_b}
and \eqref{eq:z_to_u_w}) at the vertices as well as the gluing conditions
(eqn.~\eqref{eq:gluing_cond}) for each internal edge is satisfied.  This system
leads to relations between evaluation parameters and central charges.
Generally, solutions to this system of equations are of the form
\cite{Beisert:2014qba,Kanning:2014maa,Broedel:2014pia}
\begin{align}\label{eq:CviaU}
  c_i = u_i - u_{\sigma(i)}
\end{align}
where $\sigma(i)$ is the permutation encoded by the on-shell diagram. Instead
of solving the linear system, the permutation can also be deduced graphically,
by dressing each external leg by two lines: one starting there and the other one
ending there. Drawing the lines through the diagram by turning right at each
black vertex and left at each white vertex will connect leg $i$ with its image
$\sigma(i)$, for example
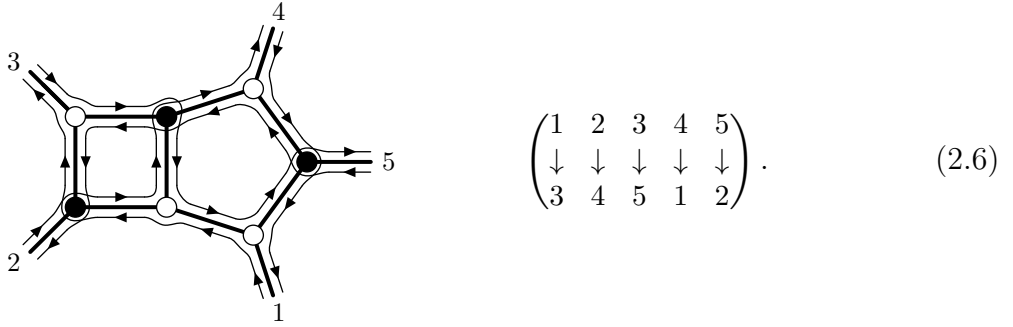
\[
  \includegraphicsboxex{Fig_5ptperm.mps}
  \qquad\qquad
  \Bigg(
  \begin{small}
  \begin{matrix}
    1 & 2 & 3 & 4 & 5\\
    \downarrow & \downarrow & \downarrow & \downarrow & \downarrow\\
    3 & 4 & 5 & 1 & 2 
  \end{matrix}
  \end{small}
  \Bigg)\,.
\]
While a unique permutation is associated to each on-shell diagram, there are
several different on-shell diagrams encoding the same permutation. Those
diagrams are related by \textit{square moves} and the \textit{merger operation}
depicted in \figref{fig:moves}. The third operation mentioned in
ref.~\cite{ArkaniHamed:2012nw}, \textit{bubble reduction}, however, modifies
the permutation. The meaning of this operation will become apparent in
\secref{sec:bubbles}.

\medskip

The main players in the $\rop$-operator formalism \cite{Chicherin:2013ora} are
the Lax operator $\lop$ and the $\rop$-operator $\rop$.  While the precise
relation between those and Yangian algebras is explained in detail in
refs.~\cite{Kanning:2014maa,Broedel:2014pia}, let us stick with the action of
the $\rop$-operator on a function defined on several copies of the on-shell
superspace, which reads
\begin{align}\label{eq:DefRop}
  \rop_{ab}(u)
  f(\lambda_a,\tilde{\lambda}_a,\tilde{\eta}_a,\lambda_b,\tilde{\lambda}_b,\tilde{\eta}_b)
  := \int_0^\infty \frac{\mathrm{d}
    z}{z^{1+u}}f(\lambda_a-z\lambda_b,\tilde{\lambda}_a,\tilde{\eta}_a,\lambda_b,\tilde{\lambda}_b+
  z \tilde{\lambda}_a,\tilde{\eta}_b+z \tilde{\eta}_a) \ .
\end{align}
In terms of these $\rop$-operators, an ansatz for a $n$-point tree-level
Yangian invariant with MHV-degree $k$ reads
\begin{align}\label{eq:AnsatzAmpl}
 \mathcal{Y} = \rop_{a_1b_1}(v_1)\ldots \rop_{a_{2n-4}b_{2n-4}}(v_{2n-4}) \Omega \ ,
\end{align}
where $\Omega$ is a product of $(n-k)$ $\dl{a_i}$'s and $k$ $\dlb{a_i}$'s which are defined as
\begin{align}\label{eq:defdelta}
  &\dl{a}:= \delta^2(\lambda_a), &&\dlb{a}:=
  \delta^{2|4}(\tilde{\lambda}_a):=\delta^{2}(\tilde{\lambda}_a)
  \delta^{4}(\tilde{\eta}_a) \ .
\end{align}
The integrations originating from $2n-4$ $\rop$-operators leave four bosonic
$\delta$-functions unintegrated, which will combine into momentum conservation
$\delta(P)$ later on. As pointed out in \secref{sec:oneloopcalc} below,
loop-level amplitudes can be constructed by applying a different number of
$\rop$-operators to a vacuum state.

The second important object in the $\rop$-operator formalism is the Lax
operator $\lop$. While rigorously defined in ref.~\cite{Chicherin:2013ora},
here it will be sufficient to note its fundamental relation with
$\rop$-operators ($u_{ab}=u_a-u_b$)
\begin{align}
  \rop_{21}(u_{12})\lop_1(u_1 + \half C_1)\lop_2(u_2 + \half C_2) &= \lop_1(u_2 + \half  C_1)\lop_2(u_1 + \half C_2) \rop_{21}(u_{12}),
  \label{eq:RLYBE}\\
  \rop_{12}(u_{12})\lop_{1}(u_{1}-\half C_1)\lop_{2}(u_{2}-\half C_2) &=
  \lop_{1}(u_{2}-\half C_1)\lop_{2}(u_{1}-\half C_2)\rop_{12}(u_{12}),
  \label{eq:RLYBEinv}
\end{align}
which is implied by the Yang--Baxter equation \cite{Broedel:2014pia}. The
equation is to be understood as an operator equation, in which the operators
$C_i$ measure the central charges on their right-hand side via
\begin{align}\label{eq:RcComm}
  &[C_a,\rop_{ab}(u)] = -u\, \rop_{ab}(u), &&[C_b,\rop_{ab}(u)] = u\,
  \rop_{ab}(u)\quad\text{and}\quad C_a\delta_{\pm a}=0\ .
\end{align}
The $n$-point monodromy matrix $\mathrm{T}_n$ is then defined as a product of
Lax-operators 
\begin{align}\label{eq:Monodromy}
  & \mathrm{T}_n := \lop_1(u_1-\sfrac{C_1}{2})\ldots \lop_n(u_n-\sfrac{C_n}{2})\,.
\end{align}
Only for certain choices of the parameters $v_1\dots v_{2n-4}$ the ansatz
eqn.~\eqref{eq:AnsatzAmpl} will yield a Yangian invariant: the condition
analogous to solving the linear system for on-shell graphs is that the ansatz
eqn.~\eqref{eq:AnsatzAmpl} has to be an eigenstate of the monodromy matrix
$\mathrm{T}_n$ defined in eqn.~\eqref{eq:Monodromy} above:
\begin{align}
  \label{eq:YangEigenValue}
  \mathrm{T}(\{u_i\})\, \mathcal{Y} = \Lambda(\{u_i\})\, \mathcal{Y} \ .
\end{align}
Commuting the monodromy matrix $\mathrm{T}_n$ through the chain of
$\rop$-operators by means of eqns.~\eqref{eq:RLYBE} and~\eqref{eq:RLYBEinv}
will fix all parameters $v_i$ in the ansatz eqn.~\eqref{eq:AnsatzAmpl} and
furthermore imply eqn.~\eqref{eq:CviaU}.

The permutation encoded in a tree-level on-shell graph is the key to expressing
a Yangian invariant in the language of $\rop$-operators
\cite{Chicherin:2013ora, Broedel:2014pia}. In order to do so, one has to
decompose the permutation into a series of successive swaps and identify the
sites to swap with the indices of $\rop$-operators $\rop_{ab}$.  Naturally,
there are many different ways of decomposing a permutation into a series of
successive swaps. For the tree-level invariants we need to consider series of
\textit{minimal length}. Only after restricting to the shortest possible
decompositions one can map a permutation to a class of on-shell diagrams (and
thus $\rop$-chains) unambiguously.  For loop-level invariants, one has to allow
for \textit{non-minimal} decompositions, which obscure the relation between
on-shell graphs and permutations.

Finally, let us comment on how to build scattering amplitudes in $\superN=4$
sYM theory from Yangian-invariant building blocks. While tree-level amplitudes
are sums of Yangian invariants themselves, for loop amplitudes only the
\emph{integrands} exhibit Yangian invariance. 
As we will see, however, the $\rop$-operator formalism provides unregulated,
integrated expressions. Since on the one hand those expressions are manifestly
Yangian invariant and on the other hand Yangian invariance is broken for loop
amplitudes due to infrared divergences, regularization needs to be responsible
for breaking Yangian invariance. 

As loops will be dealt with in the sections below, let us collect here some
facts on tree amplitudes, which have been thoroughly discussed in
refs.~\cite{Ferro:2012xw,Ferro:2013dga,Beisert:2014qba,Kanning:2014maa,Broedel:2014pia}:
in the MHV-sector \mbox{($k=2$)}, any scattering amplitude is directly related to a
single Yangian invariant.  Imposing the equations ensuring Yangian invariance
for this on-shell graphs exactly leads to the relation eqn.~\eqref{eq:CviaU}.

For other MHV-sectors ($k>2$), however, there are generically several diagrams
contributing to the scattering amplitude. While one can easily determine the
permutation associated to each of them, it is unphysical to assign different
eigenvalues $c_i$ to the same external leg. Imposing equality of all external
parameters for all contributing on-shell graphs generically forces all
eigenvalues $c_i$ to be zero\footnote{The six-point NMHV amplitude is a notable
exception.  However, while a deformed amplitude can still be defined, the
famous six-term identity \cite{Kosower:2010yk} is not valid for this deformed
amplitude.}. Thus a deformation is possible only in the MHV sector.


\section{Four-point one-loop review}
\label{sec:oneloopreview}

The four-point one-loop MHV gluon amplitude in $\mathcal{N}=4$ sYM theory was
first computed in ref.~\cite{Green1982} as the low-energy limit of the
corresponding string theory result. Later, all one-loop MHV gluon amplitudes
have been determined via unitarity in ref.~\cite{Bern1994}.  The result can be
expressed as
\begin{equation}
  \label{eq:onelBDDK}
  A_{4;2}^{(1)} = \, s\,t \, A_{4;2}^{\mathrm{tree}}I_{4}
\end{equation}
where $I_4$ is the massless box integral 
\begin{equation}
  \label{eq:Ibox}
  I_4 = \int \frac{\mathrm{d}^4 q}{q^2 (q+p_1)^2 (q+p_1+p_2)^2 (q-p_4)^2} \ ,
\end{equation}
$p_i,i=1,\dots,4$ are the null external momenta and $s = (p_1+p_2)^2,\, t =
(p_2+p_3)^2$ are Mandelstam variables. Using a supersymmetry-preserving
regulator such as dimensional reduction, the result is (see
ref.~\cite{Bern1994})
\begin{equation}
  \label{eq:onelexpl}
  A_{4;2}^{(1)} \,=\, -c_{\mathrm{\Gamma}}\,A_{4;2}^{\mathrm{tree}}
  \biggl\{
  -\frac{2}{\epsilon^2}
  \biggl[
  \biggl( \frac{\mu^2}{-s} \biggr)^{\epsilon} +
  \biggl( \frac{\mu^2}{-t} \biggr)^{\epsilon}
  \biggr] +
  \log^2 \biggl(\frac{-s}{-t}\biggr)
  + \pi^2
  \biggr\}
\end{equation}
where 
\[
c_{\mathrm{\Gamma}} = \frac{(4\pi)^\epsilon}{16 \pi^2} \,
\frac{\mathrm{\Gamma}(1+\epsilon)
  [\mathrm{\Gamma}(1-\epsilon)]^2}{\mathrm{\Gamma}(1-2\epsilon)} \ .
\]

\subsection{On-shell diagrams and all-loop BCFW}

\begin{figure}[h]
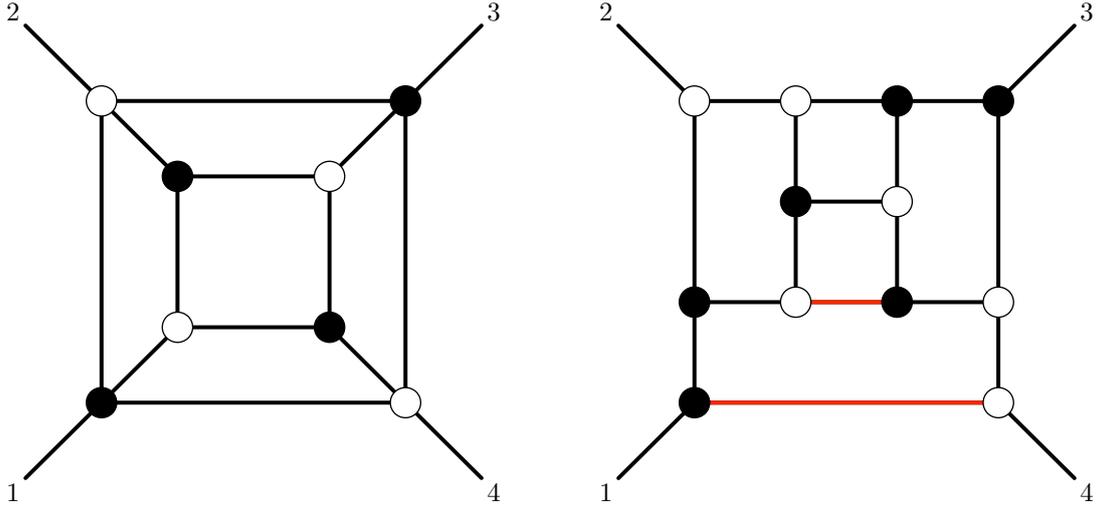

  \begin{center}
    \includegraphicsboxex[scale=1]{4pts1loopnima.mps} \hspace{1cm}
    \includegraphicsboxex[scale=1]{4pts1loopnima2.mps}
  \end{center}
  \caption{On-shell diagram corresponding to the four-point one-loop
    amplitude. }
  \label{fig:OnShellDiagram}
\end{figure}

In ref.~\cite{ArkaniHamed:2012nw}, the authors re-derived the four-point
one-loop integrand within the on-shell diagram formalism. The amplitude is
represented by the highly symmetric diagram in \figref{fig:OnShellDiagram},
which has been obtained by starting from the forward limit of the six-point
NMHV amplitude using the methods of ref.~\cite{ArkaniHamed:2010kv} and applying
several square moves and mergers afterward.  The resulting integrand is
expressed as a $\mathrm{dlog}$-form and reads
\begin{equation}
  \label{eq:nima}
  \frac{A_{4;2}^{(1)}}{A_{4;2}^{(\mathrm{tree})}} =
  \mathrm{dlog} \Bigl(  \tfrac{\alpha_1 \sprods{3}{1}}{\alpha_1 \sprods{3}{1} + \sprods{3}{4}} \Bigr) \,
  \mathrm{dlog} \Bigl( \tfrac{\alpha_2 \sprods{1}{3}}{ \alpha_2 \sprods{1}{3} + \sprods{2}{3}} \Bigr) \,
  \mathrm{dlog} \Bigl( \tfrac{\alpha_3 \sprods{1}{3}}{\alpha_3\sprods{1}{3} +  \sprods{1}{2}} \Bigr) \,
  \mathrm{dlog} \Bigl( \tfrac{\alpha_4 \sprods{3}{1}}{\alpha_4 \sprods{3}{1} + \alpha_4 \sprods{4}{1}} \Bigr) \,,
\end{equation}
where the integration variables $\alpha_i$ are BCFW-shifts not fixed by momentum
conservation and the on-shell conditions.  Alternatively, the above expression
can be rewritten in terms of an off-shell integration variable $q$, which
denotes the momentum flowing in the loop of the corresponding Feynman
diagram. The result reads
\begin{equation}
  \label{eq:onellocal}
  \frac{A_{4;2}^{(1)}}{A_{4;2}^{(\mathrm{tree})}} =
  \mathrm{dlog}\biggl( \frac{q^2}{(q-q^*)^2} \biggr) \,
  \mathrm{dlog}\biggl( \frac{(q+p_1)^2}{(q-q^*)^2} \biggr) \,
  \mathrm{dlog}\biggl( \frac{(q+p_1+p_2)^2}{(q-q^*)^2} \biggr) \,
  \mathrm{dlog}\biggl( \frac{(q-p_4)^2}{(q-q^*)^2} \biggr) \ ,
\end{equation}
where $q^* = \tfrac{\sprods{1}{2}}{\sprods{4}{2}} \lambda_4 \tilde{\lambda}_1$
is a solution of the quadruple-cut equation for the box integral \cite{Bern:2004ky}.
This integrand is equal to 
\begin{equation}
  \label{eq:integrand2}
  \mathrm{d}^4q \, \frac{s\,t}{q^2 \, (q+p_1)^2 \, (q+p_1+p_2)^2 \, (q-p_4)^2} \ ,
\end{equation}
which is exactly the integrand appearing in eqn.~\eqref{eq:onelBDDK}.
%

\subsection{Deformation of the four-point one-loop amplitude}

As pointed out in the previous subsection, the four-point one-loop amplitude
corresponds to a single on-shell diagram. Therefore it is possible to deform it
without taking care for physicality constraints arising from the compatibility
of deformations of several Yangian invariants. This was first done in
refs.~\cite{Ferro:2012xw,Ferro:2013dga}
\begin{figure}[h]
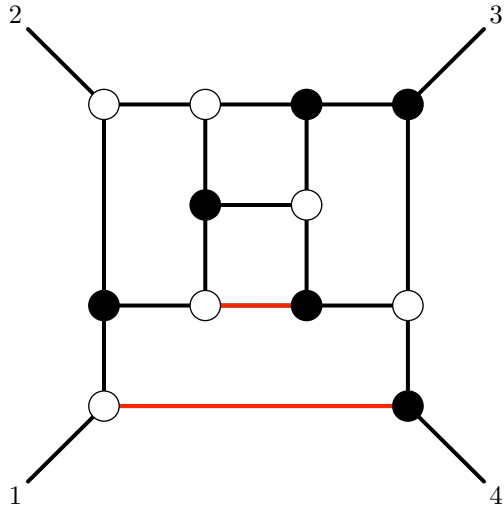

  \begin{center}
	\includegraphicsboxex[scale=1]{4pts1loopplain.mps}
  \end{center}
  \caption{The on-shell diagram corresponding to the deformed one-loop
    four-point amplitude.}
  \label{fig:OnShellDiagramJan}
\end{figure}
starting from the on-shell graph in \figref{fig:OnShellDiagramJan}. The
Yangian-invariant deformation of the amplitude reads
\begin{equation}
  \label{eq:def1l}
  \mathcal{A}_{4;2}^{(1)} \, = \,
  s\,t\,A_{4;2}^{\mathrm{tree}}\, \tilde{I}_{4}({a_i};s,t) \ ,
\end{equation}
where $\tilde{I}_{4}({a_i};s,t)$ is a box integral with the propagators raised
to arbitrary complex powers,
\begin{equation}
  \label{eq:defbox}
  \tilde{I}_{4}({a_i};s,t) \, = \,
  \int \mathrm{d}^4 q\,
  \frac{1}{[(q)^2]^{1+a_1}
    [(q + p_1 )^2]^{1+a_2}
    [(q + p_1 + p_2 )^2]^{1+a_3}
    [(q - p_4 )^2]^{1+a_4} } \ .
\end{equation}
This form of the integrand is reminiscent of analytic regularization.

Imposing Yangian invariance for the on-shell diagram
implies that $\sum_i a_i = 0$ \cite{Beisert:2014qba}. The explicit computation
of this integral (with the Yangian invariance condition enforced) is subtle.
If however one does not enforce Yangian invariance, the computation can lead to a
finite answer. With a specific choice of external central charges $c_1 = c_2 =
-c_3 = -c_4 = 4 \epsilon$ (equivalent to choose all $a_i = \epsilon$), the
explicit computation leads to \cite{Ferro:2012xw,Ferro:2013dga}
\begin{equation}
  \label{eq:1ldefres}
  A_{4;2}^{(1)} \, = \, A_{4;2}^{(\mathrm{tree})}
  \biggl( \frac{\cprods{3}{4}}{\cprods{1}{2}} \biggr)^{4\epsilon}
  \biggl[
  \frac{1}{\epsilon^2} \biggl(\frac{s}{t}\biggr)^{-2\epsilon}
  -\frac{1}{2} \biggl(\log \frac{s}{t}\biggr)^2
  -\frac{7 \pi^2}{6} + \mathcal{O}(\epsilon)
  \biggr]\,,
\end{equation}
which bears a striking resemblance with the dimensionally regulated version in
eqn.~\eqref{eq:onelexpl}.

As pointed out before, Yangian invariance for the on-shell diagram is
equivalent to demanding $\sum_i a_i = 0$. This condition makes the computation
of the integral less straightforward. The result seems to be a distribution
with support on the surface $a_1 - a_3=0, \, a_1 + a_2 = 0$
\cite{Beisert:2014qba}:
\begin{equation}
  \label{eq:onelniklas}
  \begin{aligned}
    A_{4;2}^{(1)} &= \, s \,t \, A_{4;2}^{(\mathrm{tree})} \,f(a_1,a_2,a_3) \ ,
    \quad \text{where} \\
    f(a_1,a_2,a_3) &= - \delta(a_1+a_2) \delta(a_2+a_3) \frac{1}{s\,t}
    \biggl(\frac{t}{s}\biggr)^{a_1} \frac{\sin (\pi a_1)}{a_1} \ .
  \end{aligned}
\end{equation}
That is, for almost all deformations, the integral vanishes.


\section{Bubbles}
\label{sec:bubbles}

Among the configurations appearing in on-shell diagrams, the bubble takes a
special r\^ole. It reflects the double lines: 
\[
\includegraphicsboxex{bubbleperm2.mps}\ .
\]
In the $\rop$-operator language, the above diagram will be produced by the Yangian invariant: 
\begin{equation}
\label{eq:bubble1}
\rop_{ab}(u_{ab})\rop_{ab}(u_{ba})\dl{a}\dlb{b}\,.
\end{equation}
The very same trivial permutation, however, is represented by
$\rop_{ba}(u_{ba})\rop_{ba}(u_{ab})\dlb{a}\dl{b}$, which corresponds to the
following diagram:
\[
\includegraphicsboxex{bubbleperm.mps}\,.
\]
Employing the definition of the $\rop$-operator eqn.~\eqref{eq:DefRop} in the
first case eqn.~\eqref{eq:bubble1} leads to
\[
\int\frac{\mathrm{d} z_1}{z_1^{1+u_{ab}}}\int\frac{\mathrm{d}
  z_2}{z_2^{1+u_{ba}}}\delta^2(\lambda_a+(z_1+z_2)\lambda_b)\delta^2(\tilde{\lambda}_b-(z_1+z_2)\tilde{\lambda}_a)\,.
\]
Changing variables to $w=z_1+z_2$ and $z=z_1-z_2$, performing the integration
over $w$ and substituting the remaining variable then leads to
\begin{equation}
\label{eqn:basicintegral}
\int\frac{\mathrm{d}z}{z^{1+u_{ab}}(D-z)^{1-u_{ab}}}
\,\delta(\sprods{a}{b})\delta^2(\tilde{\lambda}_b-\frac{\lambda^1_a}{\lambda^1_b}\tilde{\lambda}_a)\,,
\end{equation}
%
where $D$ is some function of the external kinematics. The above integral
exhibits several important features:
\begin{itemize}
\item the kinematics is constrained by $\delta(\sprods{a}{b})$. While this
  renders the kinematics special in the situation of an isolated bubble, it is
  nothing to worry about, if the bubble is part of a larger on-shell diagram.
\item the integration over $z$ means that momentum conservation does not
  constrain the kinematics completely: the parameter $z$ measures, which part of
  the momentum flows along the upper and which part along the lower line in the
  bubble.
\item the integration variable $z$ does not appear in the arguments of the
  $\delta$-function: this allows to consider the integration separately. In
  fact, this is exactly the situation described by the \textit{bubble deletion}
  operation introduced in ref.~\cite{ArkaniHamed:2012nw} and mentioned already
  in \secref{sec:ampl}: one can replace the bubble by a line after factoring out
  an integration. The link to the integral is provided by the function $D$ of
  the external kinematics.
  \[
  \includegraphicsboxex{bubbledeletion1.mps}
  \quad\rightarrow\quad
  \includegraphicsboxex{bubbledeletion2.mps}
  \]
  Note, however, that the operation does \textit{not} preserve the permutation
  encoded by the on-shell diagram and thus the Yangian structure, e.g. the flow
  of the central charges and evaluation parameters is modified by deleting a
  bubble.
\end{itemize}

\begin{figure}[<+htpb+>]
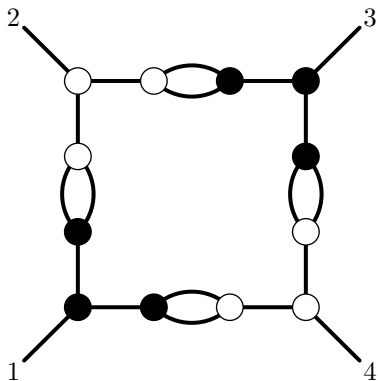

  \begin{center}
    \includegraphicsboxex{bubblegraph.mps}\,.
  \end{center}
  \caption{Yet another representation of the four-point one-loop amplitude.}
  \label{fig:bubblegraph}
\end{figure}

Below we will see that the integral eqn.~\eqref{eqn:basicintegral} encoded in
the bubble is important for investigating deformations of one-loop amplitudes.
The close relation of bubbles with loop diagrams becomes apparent in particular
by considering the diagram in \figref{fig:bubblegraph}, which is another form
of representing the four-point one-loop amplitude. Using the square moves and
mergers, it can be transformed
into the diagrams in \figref{fig:OnShellDiagram}. It is not difficult to see
that all these diagrams encode the trivial permutation
\[
\begin{permutation}
  1 & 2 & 3 & 4\\
  \downarrow & \downarrow & \downarrow & \downarrow\\
  1 & 2 & 3 & 4
\end{permutation}.
\]
%

\paragraph{Contour of integration}
Leaving the Grassmann variables aside for a moment, the $\rop$-operator as defined in
eqn.~\eqref{eq:DefRop} reads
\begin{equation}
  \label{eq:rdef1}
  R_{ij}(u)\cdot F(\{\lambda,\tilde{\lambda},\tilde{\eta}\})
  := \frac{i}{2 \sin (\pi u)} \int_{\mathrm{\Gamma}}  \frac{\mathrm{d}z}{z^{1+u}} \, F(\lambda_i - z \lambda_j,\tilde{\lambda}_j+z \tilde{\lambda}_i) \ ,
\end{equation}
Here the contour $\mathrm{\Gamma}$ encircles the branch cut of the integrand
(chosen to lie on the positive real axis) going around zero; this open contour
is known as the Hankel contour.
%
%
Already here we encounter a problem, since the Hankel contour does not take into
account possible branch cuts from the integrated function $F$ itself. We are
going to see below that this is indeed a problem for defining an integration
contour for loop amplitudes.  However, even for the action of only a single
$\rop$-operator the situation is not completely clear. One could guess that the
Hankel contour is the correct choice in general, since one acts always on
single-valued functions; this is however a bit misleading.  In order to see
this, we study the integral over $z$ in eqn.~\eqref{eqn:basicintegral}. Even
though one could argue that this integral represents a degenerate situation (a
deformed on-shell diagram with one single bubble), we will see that the same
integral appears in the deformed four-point one-loop amplitude below.

\begin{figure}
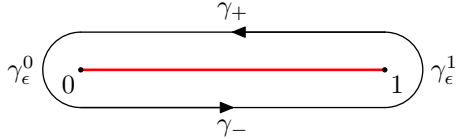

  \centering
  \includegraphicsboxex{branchcont.mps}
  \caption{Contour around the branch cut in the $\zeta$ plane}
  \label{fig:branchcontour}
\end{figure}

The question now is how to choose the integration contour; a naive guess would
be simply to generalize the Hankel contour to the \emph{closed} contour
encircling the branch cut, so that the integral picks up the discontinuity of
the integrand along the cut.  With the change of variables $z = D \zeta$, the
integral becomes proportional to (here $\alpha=-u_{ab}$)
\begin{equation}
  \label{eq:int2}
  \oint_{\Gamma} \frac{\mathrm{d}\zeta}{\zeta^{1-\alpha} (1 - \zeta)^{1+\alpha}} \ ,
\end{equation}
where $\Gamma$ is the contour encircling the branch cut between $0$ and $1$.
Specifically, $\Gamma = \gamma^0_\epsilon\cup \gamma_+ \cup
\gamma^1_{\epsilon}\cup \gamma_- $, where (see \figref{fig:branchcontour})
\begin{itemize}
\item $\gamma^0_\epsilon,\,\gamma^1_\epsilon$ are semicircles around the two
  branch points $0,1$ of radius $\epsilon$;
\item $\gamma_\pm$ are line segments from $0\pm i\epsilon$ to $1 \pm i\epsilon$
  (with the correct orientation).
\end{itemize}
The integrand has no pole at infinity, therefore the evaluation via residue
theorem gives zero. Notice, however, that the vanishing of the integral is
nontrivial when we consider the single contributions, due to the fact that for
$\mathrm{Re}(\alpha)>0$ the integral around $\gamma_\epsilon^1$ and the
contributions from $\gamma_+\cup \gamma_-$ diverge (and similarly for
$\mathrm{Re}(\alpha)<0$ with $\gamma_\epsilon^0$)~\footnote{ In this case we can
  simply evaluate directly the real integral $\int_0^1
  \frac{\mathrm{d}\zeta}{\zeta^{1-\alpha} (1 - \zeta)^{1+\alpha}} =
  \frac{z^\alpha}{(1-z)^\alpha}\bigr\vert_0^1$, which of course diverges unless
  $\mathrm{Re}(\alpha) = 0$. }. Obviously, even defining the integral purely as
the discontinuity around the branch cut (i.e. consider only the contributions
along $\gamma_+\cup \gamma_-$) leads to an ill-defined answer, therefore it
seems nontrivial to define an open contour generalizing the Hankel contour.

Notice that the situation considered in the present paragraph is qualitatively
different from the one considered in ref.~\cite{Ferro:2012xw,Ferro:2013dga};
since in these references the scaling of the deformed integrands depends on the
deformation parameters. In our simple case eq.~\eqref{eq:int2}, the fact that
the exponents of the denominator of the integrand sum to zero implies that
$z=\infty$ is a regular point and thus the branch-cut can be chosen to lie on
the real axis between $0$ and $1$. However, if the exponents sum to a noninteger
value, then the integrand would behave as $z^{2-\beta}$ for some
$\beta\in\mathbb{C}$ as $|z|\to\infty$. The branch-cut would then lie along
$\mathbb{R}\backslash (0,1)$. In that case, the simple integral we consider
could be evaluated exactly as a real integral between $0$ and $1$, and would
lead to a finite and well-defined answer in a definite region of the space of
deformation parameters.

This discussion is done to stress the fact that when considering tree-level
amplitudes the contour can be chosen a priori due to the knowledge that all the
integrals will be localized on the support of delta functions, but the situation
is not as clear when considering integrals beyond tree-level.

As stated at the beginning of this paragraph, we will see that integrals of the
form in eqn.~\eqref{eq:int2} appear in the computation of the deformed
four-point one-loop amplitude. Specifically, eqn.~\eqref{eq:deformed1l} from the
following section reads
%
\begin{equation}
  \label{eq:1lnima}
  \begin{aligned}
    \int &\frac{\mathrm{d}z_1\, \mathrm{d}z_2\, \mathrm{d}z_3\, \mathrm{d}z_4 }{z_1^{1-u_{14}}\, z_2^{1-u_{21}}\, z_3^{1-u_{32}} \,z_4^{1-u_{43}}} \times \\
    &\quad\times \Bigl[ \tfrac{ \sprods{3}{4} }{ \sprods{3}{4} - z_1
      \sprods{3}{1} } \Bigr]^{1+u_{14}} \Bigl[ \tfrac{ \sprods{2}{3} }{
      \sprods{2}{3} - z_2 \sprods{1}{3} } \Bigr]^{1+u_{21}} \Bigl[ \tfrac{
      \sprods{1}{2} }{ \sprods{1}{2} - z_3 \sprods{1}{3} } \Bigr]^{1+u_{32}}
    \Bigl[ \tfrac{ \sprods{4}{1} }{ \sprods{4}{1} - z_4 \sprods{3}{1} }
    \Bigr]^{1+u_{43}}
  \end{aligned}
\end{equation}
The four integrals are completely independent, so we can consider one at a time,
for example
\begin{equation}
  \label{eq:int1}
  \int \frac{\mathrm{d}z}{z^{1-a} (\sprods{4}{1} - z \sprods{3}{1})^{1+a}}\ ,
\end{equation}
where the integrand has a single branch cut between $z=0$ and
$z=z^*:=\frac{\sprods{4}{1}}{\sprods{3}{1}}$. Each of the integrals in
eqn.~\eqref{eq:1lnima} is of the type of eqn.~\eqref{eq:int2}.

Summing up, it seems then that the deformation leads to well-defined integrands,
but the contours of integration are somewhat tricky to define. This fact is
immaterial at tree level, since all the integrals are localized on the support
of the delta functions. However, at loop level this ambiguity should be resolved
in order to attempt to explicitly compute the result.

There are various possible solutions to the contour problem.  It is possible to
work at the level of the integrand and find a coordinate transformation that
leads to a deformed box integral, and then define the usual domain of
integration in the loop momentum space, analogous to the computation of
refs.\cite{Ferro:2012xw,Ferro:2013dga}; this will be the approach implicitly
followed in the following sections.  It is also conceivable to define a
different contour that crosses the branch cuts of the integrands. The knowledge
\emph{a priori} of the position of the branch cuts of the integrand is then
crucial to define such a contour \cite{StaudacherAmpl}.



\section{Four-point one-loop calculation}
\label{sec:oneloopcalc}

In this section we will discuss the derivation of the deformed one-loop
four-point amplitude in the language of $\rop$-operators. We will investigate
the eigenstates of the monodromy that correspond to \figref{fig:OnShellDiagram}
and \figref{fig:OnShellDiagramJan} under the dictionary given in
\cite{Broedel:2014pia}. The evaluation of these eigenstates yields integrals
which we will map to the deformed box-integral of eqn.~\eqref{eq:defbox}.

\paragraph{Tree level.}
Let us start by considering the four-point tree-level amplitude. As shown in
\cite{Broedel:2014pia}, the four-point tree-level Yangian invariant can be
represented as an eigenfunction of the monodromy matrix of the form
\begin{align}\label{eq:tree4}
 \mathcal{A}^{(0)}_{4;2} = \rop_{23}(u_{32})\rop_{34}(u_{42})\rop_{12}(u_{31})\rop_{23}(u_{41})
\Omega_{++--}.
\end{align}
It has eigenvalue $(u_1+\half)(u_2+\half)(u_3-\half)(u_4-\half)$.  Its central
charges are readily computed to be
\begin{align}
&c_1 = u_1 - u_3,
&&c_2 = u_2 - u_4,
&&c_3 = u_3 - u_1,
&&c_4 = u_4 - u_2,
\end{align}
which, using eqn.~\eqref{eq:CviaU},  can be directly translated into the permutation
that defines this tree-level Yangian invariant. It is simply a shift by two:
\[
    \begin{permutation}
     1 & 2 & 3 & 4\\
     \downarrow & \downarrow & \downarrow & \downarrow\\
     3 & 4 & 1 & 2
    \end{permutation}.
\]
Employing the definition of the $\rop$-operator eqn.~\eqref{eq:DefRop} and the vacuum
\eqref{eq:defdelta} we can evaluate eqn.~\eqref{eq:tree4} explicitly in terms of
spinor-helicity variables
\begin{align}\label{eq:defMHV4}
  \mathcal{A}^{(0)}_{4;2}(\{u_i\}) = \frac{\delta^4(\sum_i p_i)\delta^{8}(\sum_i
    \lambda_i\eta_i)}{\langle
    12\rangle^{1+u_{32}}\langle23\rangle^{1+u_{43}}\langle
    34\rangle^{1+u_{14}}\langle 14 \rangle^{1+u_{21}}},
\end{align}
For $u_{ij}=0$ this expression reduces to the well-known MHV tree-level amplitude for $\superN=4$ sYM theory.

\paragraph{One-loop eigenstate.} 

As discussed in \secref{sec:oneloopreview}, the permutation corresponding to the
one-loop diagrams in \figref{fig:OnShellDiagram}  and \figref{fig:OnShellDiagramJan}  is the trivial one.
At tree level, the trivial permutation clearly corresponds to the ground state
$\Omega$. Nevertheless, we can generate non-trivial Yangian invariants that are
associated with these diagrams.

In ref.~\cite{Broedel:2014pia} the explicit map between $\rop$-operators and
on-shell diagram has been discussed. It is easy to check that the following two states
\begin{align}\label{eq:Rop1L4P}
\mathcal{A}_{4;2}^{(1)}=
\rop_{41}(u_{14})\rop_{21}(u_{21})\rop_{23}(u_{32})\rop_{43}(u_{43})
\rop_{23}(u_{23})\rop_{34}(u_{13})\rop_{12}(u_{24})\rop_{23}(u_{14})
\Omega_{++--},\\
\mathcal{B}_{4;2}^{(1)}=
\rop_{14}(u_{14})\rop_{21}(u_{24})\rop_{23}(u_{32})\rop_{43}(u_{13})
\rop_{23}(u_{23})\rop_{34}(u_{43})\rop_{12}(u_{21})\rop_{23}(u_{41})
\Omega_{++--}\label{eq:Rop1L4Pb}
\end{align}
exactly give rise to the two on-shell diagrams in \figref{fig:OnShellDiagram}
and \figref{fig:OnShellDiagramJan} respectively. Notice that the tree-level
invariant eqn.~\eqref{eq:tree4} is represented in the above expressions
manifestly by the four rightmost $\rop$-operators (up to a redefinition of the
spectral parameters).

It can be readily shown that both states are eigenstates of the monodromy matrix
with eigenvalues~\footnote{Details are given in Appendix \ref{app:EigenV}.}
\begin{align}\label{eq:EigenVLoop4}
\mathrm{T}\, \mathcal{A}_{4;2}^{(1)} = (u_1+\half)(u_2-\half)(u_3+\half)(u_4-\half)\mathcal{A}_{4;2}^{(1)},\\
\mathrm{T}\, \mathcal{B}_{4;2}^{(1)} = (u_1-\half)(u_2-\half)(u_3+\half)(u_4+\half)\mathcal{B}_{4;2}^{(1)}.
\end{align}
The eigenvalues are simply related by interchanging $u_1$ and $u_4$.
Furthermore, it is quickly seen that the central charges vanish
\begin{align}
c_i = 0,
\end{align}
indicating that eqns.~\eqref{eq:Rop1L4P} and \eqref{eq:Rop1L4Pb} indeed
correspond to the trivial permutation.

In other words, we have identified Yangian invariants that correspond to the
four-point one-loop amplitude.  At this point we would like to note that both
$\mathcal{A}_{4;2}^{(1)}$ and $\mathcal{B}_{4;2}^{(1)}$ are valid deformations
of the one-loop amplitude.  
If these integrals are well-defined, the fact that they have different
eigenvalues implies that they either need to be inequivalent or trivial.
Notice furthermore, that apart from applying the parity flip to particles one
and four, we could also have applied a flip to any other set of neighboring
particles, leading to different deformations.  We will come back to this
important point at the end of this section.

\paragraph{Integral $\mathcal{A}$.}

Having established the relation between the two one-loop eigenstates and their
on-shell diagrams, let us evaluate the loop integrals they generate. We
will show that both, eqns.~\eqref{eq:Rop1L4P} and \eqref{eq:Rop1L4Pb}, give rise to
the integral eqn.~\eqref{eq:def1l}. We will first treat eqn.~\eqref{eq:Rop1L4P} in
great detail and then briefly discuss the derivation of the integral
corresponding to eqn.~\eqref{eq:Rop1L4Pb} afterward.

The Yangian invariant eqn.~\eqref{eq:Rop1L4P} naturally contains eight integrations.
The four rightmost $\rop$-operators generate the tree-level Yangian invariant and
consequently it is convenient to first perform these integrations. This
will remove four $\delta$-functions from the vacuum and leaves us with an
expression of the form
\begin{align}\label{eq:4ptIntegral}
  \mathcal{A}_{4;2}^{(1)} = (-1)^{1+u_{41}}\mathcal{A}_{4;2}^{(0)}(\{0\})\, I_4(\{u_i\}),
\end{align}
where the integral $I_4(\{u_i\})$ is four-dimensional and depends on the
integration variables $z_1,z_2,z_3,z_4$ from the remaining $\rop$-operators.
We would like to stress that the factorization in \eqref{eq:4ptIntegral} can only be done after
the application of \textit{all} R-operators. At the algebraic level, there does not seem to be a
way to factor the amplitude into the tree-level amplitude times an integral.
Equation \eqref{eq:4ptIntegral} can be derived straightforwardly by applying eqn.~\eqref{eq:DefRop}
to the four-point result eqn.~\eqref{eq:defMHV4} . This yields for the integral $I_4(\{u_i\})$ (cf. \cite{ArkaniHamed:2010kv})
\begin{align}
\int 
  \frac{\frac{\langle34\rangle}{\langle31\rangle}\,\mathrm{d}z_1}{z_1^{1-u_{14}}\!
    \left[\!\frac{\langle34\rangle}{\langle31\rangle}- z_1\right]^{1+u_{14}}}
  \frac{\frac{\langle23\rangle}{\langle13\rangle}\,\mathrm{d}z_2}{z_2^{1-u_{21}}\!
    \left[\!\frac{\langle23\rangle}{\langle13\rangle}- z_2\right]^{1+u_{21}}}
  \frac{\frac{\langle12\rangle}{\langle13\rangle}\,\mathrm{d}z_3}{z_3^{1-u_{32}}\!
    \left[\!\frac{\langle12\rangle}{\langle13\rangle}- z_3\right]^{1+u_{32}}}
  \frac{\frac{\langle41\rangle}{\langle31\rangle}\,\mathrm{d}z_4}{z_4^{1-u_{43}}\!
    \left[\!\frac{\langle41\rangle}{\langle31\rangle}- z_4\right]^{1+u_{43}}}\,.
\label{eq:deformed1l}
\end{align}
This integral should match the deformed box integral eqn.~\eqref{eq:defbox}. We
see that the integrations of the $\rop$-operator will play the role of the loop
momentum. 
%
%
In order to match this to eqn.~\eqref{eq:defbox}, we have to define what the
loop momentum is in our integral. The BCFW recursion relation for loops (or the
forward limit)  provides a natural candidate: it can be obtained by summing the
momenta flowing along the red lines in \figref{fig:OnShellDiagram} and
\figref{fig:OnShellDiagramJan}. The momentum propagating along the various BCFW
bridges clearly is a function of the integration parameters $z_i$ corresponding
to the $\rop$-operators that generate the loop integral (the left-most four
R-operators).  Let us define the shifted momenta \cite{ArkaniHamed:2010kv}
\begin{align}
&\lambda_{\hat{1}} = \lambda_{1}, &&
\tilde{\lambda}_{\hat{1}} = \tilde{\lambda}_{1} + z_2 \tilde{\lambda}_{2} + z_1\tilde{\lambda}_{4}, \nln
&\lambda_{\hat{2}} = \lambda_{2} - z_2 \lambda_1 - z_3 \lambda_3, 
&&\tilde{\lambda}_{\hat{2}} = \tilde{\lambda}_{2},\nln
&\lambda_{\hat{3}} = \lambda_{3}, &&
\tilde{\lambda}_{\hat{3}} = \tilde{\lambda}_{3} + z_3 \tilde{\lambda}_{2} + z_4\tilde{\lambda}_{4}, \nln
&\lambda_{\hat{2}} = \lambda_{2} - z_1 \lambda_1 - z_4 \lambda_3, 
&&\tilde{\lambda}_{\hat{2}} = \tilde{\lambda}_{2}.
\end{align}
These expressions can be easily reproduced by acting with the relevant
$\rop$-operators in eqn.~\eqref{eq:Rop1L4P} on the momentum and removing the
measure and integration, i.e. by only considering the shift part from eqn.
\eqref{eq:DefRop}. Let us denote such a shift corresponding to the R-operator
$\rop_{ab}$ by $\mathrm{S}_{ab}(z)$, then
\begin{align}
&\lambda_{\hat{a}} = 
\mathrm{S}_{41}(z_1)\mathrm{S}_{21}(z_2)\mathrm{S}_{23}(z_3)\mathrm{S}_{43}(z_4)\lambda_a,
&&\tilde{\lambda}_{\hat{a}} = 
\mathrm{S}_{41}(z_1)\mathrm{S}_{21}(z_2)\mathrm{S}_{23}(z_3)\mathrm{S}_{43}(z_4)\tilde{\lambda}_a.
\end{align}
By momentum conservation this results in the following natural expression for
the loop momentum
\begin{align}
  q := \frac{\langle\hat{1}\hat{2}\rangle}{\langle\hat{2}\hat{4}\rangle}
  \lambda_{\hat{4}}\tilde{\lambda}_{\hat{1}} + z_1
  \lambda_{1}\tilde{\lambda}_{4}.
\end{align}
This defines the coordinate transformation between the shift variables $z_i$
and the loop momentum. Given this transformation, it is now straightforward to
show that 
\begin{equation}\label{eq:BoxInt1}
  I_{4}({a_i};s,t) \, = \,
  \int \mathrm{d}^4 q\,
\frac{s \, t}
  {[(q)^2]^{1+u_{41}}
    [(q + p_1 )^2]^{1+u_{12}}
    [(q + p_1 + p_2 )^2]^{1+u_{23}}
    [(q - p_4 )^2]^{1+u_{34}} } \ ,
\end{equation}
which exactly matches eqn.~\eqref{eq:defbox}.

\paragraph{Integral $\mathcal{B}$.} Let us now briefly indicate what happens for
the eigenstate eqn.~\eqref{eq:Rop1L4Pb}. The integrand is again of similar type
as the integrand of $\mathcal{A}$.  Next, we define shifted spinor helicity
variables according to the $\rop$-operators associated to
eqn.~\eqref{eq:Rop1L4Pb} as
\begin{align}
&\lambda_{\check{a}} = 
\mathrm{S}_{14}(z_1)\mathrm{S}_{21}(z_2)\mathrm{S}_{23}(z_3)\mathrm{S}_{43}(z_4)\lambda_a,
&&
\tilde{\lambda}_{\check{a}} = 
\mathrm{S}_{14}(z_1)\mathrm{S}_{21}(z_2)\mathrm{S}_{23}(z_3)\mathrm{S}_{43}(z_4)\tilde{\lambda}_a,
\end{align}
and the corresponding expression for the loop momentum is this time given by
\begin{align}
  q := \frac{\langle\check{1}\check{2}\rangle}{\langle\check{2}\check{4}\rangle}
  \lambda_{\check{4}}\tilde{\lambda}_{\check{1}} - z_1
  \lambda_{4}\tilde{\lambda}_{1}.
\end{align}
Remarkably, this coordinate transformation maps the integral corresponding to
eqn.~\eqref{eq:Rop1L4Pb} to eqn.~\eqref{eq:BoxInt1} as well. In other words,
eqns.~\eqref{eq:Rop1L4P} and \eqref{eq:Rop1L4Pb} correspond to the same
function in momentum space but originate from different Yangian invariants.

\paragraph{Summary.} We have shown that both eqns.~\eqref{eq:Rop1L4P} and
\eqref{eq:Rop1L4Pb} give rise to the same Yangian-deformed box integral. This
means that the deformed amplitude is the same for both on-shell
diagrams \figref{fig:OnShellDiagram} and \figref{fig:OnShellDiagramJan}.
However, as pointed out in eqn.~\eqref{eq:EigenVLoop4}, both states have
different eigenvalues under the monodromy matrix. This can only happen if
either the corresponding states vanish, are ill-defined or if their eigenvalues
coincide.

For these two states, the eigenvalues agree exactly when $u_1=u_4$. However, as
was indicated in the beginning of this section, we could have chosen any two
adjacent particles and construct the analogue of $\mathcal{B}^{(1)}_{4;2}$. In
other words, in order for all these eigenvalues to coincide, we need that
$u_i=u_{i+1}$. This can only be accomplished when the deformation is trivial,
\textit{i.e.} $u_{ij}=0$, which renders the integral the usual, unregulated
box integral. 

Indeed in \cite{Beisert:2014qba} it was shown that the integration of
\eqref{eq:BoxInt1} on $\mathbb{R}^4$ leads to a vanishing result for generic
values of the deformation parameters.
Furthermore, in \cite{Beisert:2014qba} the integral seems to have singular
support, but we find that even in those particular cases the integral will
either vanish or be ill-defined (\textit{i.e. divergent}).  We find that the
unregulated integral follows from a manifestly Yangian invariant procedure.
However, due to the fact that different Yangian invariants give rise to the same
integral, this Yangian invariant seems to be either 0 or divergent.

Finally, as remarked in \cite{Broedel:2014pia}, the eigenvalue of a Yangian
invariant corresponds to the hypercharge and its Yangian partners. It is easy to
see that the hypercharge of the invariants will start to differ at the second
Yangian level only.  In turn, this means that our conundrum is closely related
to the fact that we extended our algebra not only by the central charge
operators $\alg{C}$ but as well by the hypercharge $\alg{B}$.

It is also conceivable that the manipulations of the two expressions in
eq.~\eqref{eq:Rop1L4P} and \eqref{eq:Rop1L4Pb} are valid at the level of the
\emph{integral} and therefore we should carefully consider the transformation of
the contours under the change of variables, as well. However, due to the lack of
a precise definition of the contour for the general action of $\rop$-operators,
we are not able to make any conclusive statement yet.


\section{Five-point one-loop amplitude}
\label{sec:onshell}

In this section we discuss the five-point one-loop amplitude. This amplitude is
obtained from three on-shell diagrams. We will show that each of these diagrams
corresponds to Yangian invariants. However, they have different central charges.
It turns out that this implies that they can only be added if the deformation is
trivial.

\begin{figure}
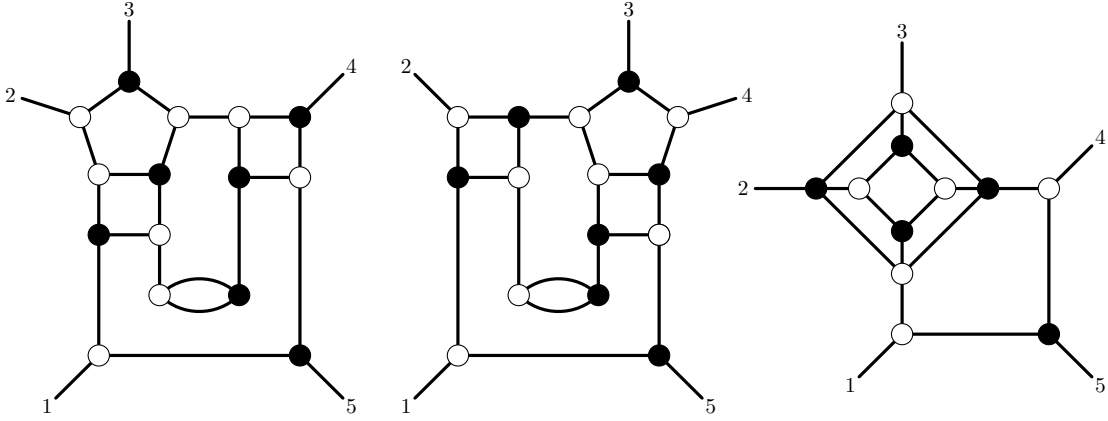

  \centering
  \includegraphicsboxex[scale=0.8]{5pts1loopa.mps}
  \quad
  \includegraphicsboxex[scale=0.8]{5pts1loopb.mps}
  \hspace{-.5cm}
  \includegraphicsboxex[scale=0.8]{5pts1loopc.mps}
  \caption{Three channels contributing to the five-point one-loop amplitude.}
  \label{fig:5pt1loopchannels}
\end{figure}

Following the BCFW-recursion relation for loop amplitudes, we find that the
diagrams in \figref{fig:5pt1loopchannels} contribute. The first two diagrams
correspond to a forward limit of a seven-point amplitude. The last diagram is
an inverse soft limit of the four-point one-loop amplitude discussed in the
previous section. The permutations associated to each of the diagrams are
\[
\label{eq:Perm5pt}
    \begin{permutation}
     1 & 2 & 3 & 4 & 5\\
     \downarrow & \downarrow & \downarrow & \downarrow & \downarrow\\
     3 & 1 & 2 & 4 & 5 
    \end{permutation}\ , \;
    \begin{permutation}
     1 & 2 & 3 & 4 & 5\\
     \downarrow & \downarrow & \downarrow & \downarrow & \downarrow\\
     4 & 2 & 1 & 3 & 5 
    \end{permutation}
    \quad\text{and}\quad
    \begin{permutation}
     1 & 2 & 3 & 4 & 5\\
     \downarrow & \downarrow & \downarrow & \downarrow & \downarrow\\
     5 & 2 & 3 & 1 & 4 
    \end{permutation},
\]
respectively, which can be quickly derived using the double-line notation
introduced in ref.~\cite{Beisert:2014qba}.

Let us now spell out the chains of R-operators generating these on-shell
diagrams. Define the following three states
\begin{align}
\mathcal{A}^{(1)}_{5;2} =&
\rop_{51}(u_{15}) \rop_{45}(u_{41}) \rop_{54}(u_{54}) 
\rop_{12}(u_{52}) \rop_{51}(u_{24}) \rop_{34}(u_{13}) \nln
&\qquad \rop_{15}(u_{34}) \rop_{45}(u_{23}) \rop_{42}(u_{52}) \rop_{32}(u_{21})
\Omega_{+-++-}\label{eq:loop5a}\\
\mathcal{B}^{(1)}_{5;2} =&
\rop_{51}(u_{15}) \rop_{12}(u_{25}) \rop_{34}(u_{43}) \rop_{45}(u_{13})
\rop_{51}(u_{23}) \rop_{21}(u_{24}) \nln
&\qquad \rop_{15}(u_{25}) \rop_{41}(u_{31}) \rop_{31}(u_{14}) \rop_{21}(u_{45})
\Omega_{-+++-}\label{eq:loop5b}\\
\mathcal{C}^{(1)}_{5;2} =&
\rop_{51}(u_{15}) \rop_{54}(u_{41}) \rop_{41}(u_{54}) \rop_{23}(u_{32})
\rop_{43}(u_{43}) \rop_{23}(u_{23}) \nln
&\qquad \rop_{34}(u_{53})
\rop_{13}(u_{54}) \rop_{23}(u_{52}) \rop_{23}(u_{52})
\Omega_{++--+}.\label{eq:loop5c}
\end{align}
It can be shown that they indeed are eigenstates of the monodromy matrix in a
way similar to the discussion in Appendix \ref{app:EigenV}. Furthermore, it
follows directly that these three Yangian invariants generate the on-shell
diagrams listed in \figref{fig:5pt1loopchannels}. 

In order to study the permutation corresponding to these Yangian invariants, we
compute the central charges
\begin{align}
&c_\mathcal{A} = \{u_1-u_3,u_2-u_1,u_3-u_2,0,0 \},\nln
&c_\mathcal{B} = \{u_1-u_4,0,u_3-u_1,u_4-u_3,0 \},\\
&c_\mathcal{C} = \{u_1-u_5,0,0,u_4-u_1,u_5-u_4 \}\nonumber.
\end{align}
According to eqn.~\eqref{eq:CviaU} we find that these agree with the
permutations listed in eqn.~\eqref{eq:Perm5pt}. 

Similar to NMHV amplitudes at tree level, we need to combine several terms to
form the full amplitude. Obviously we can only add terms that belong to the
same eigenspace, \textit{i.e.} the central charges must agree. This imposes
conditions on the evaluation parameters $u_i$ and it is readily checked that
implies that $u_{ij}=0$. In other words, there is \textit{no} deformed
five-point one-loop amplitude if one insists on compatibility of the Yangian
invariance of the individual terms.

\section{Conclusions}
\label{sec:conclusions}

In this paper we have constructed and evaluated one-loop diagrams in the
language of R-operators. We find that for four points, the \textit{same} loop
integral can be recovered from different on-shell diagrams. However, since
formally these two diagrams correspond to different Yangian invariants, it
turns out that the integrals either evaluate to zero or diverge.

Except for the four-point one-loop amplitude, there is no consistent deformation
for loop amplitudes. The reason is the same, which ruled out the deformed tree
amplitudes: if there are several Yangian invariants contributing to a
loop-integrand, demanding the same central charges for the external legs of each
diagram constrains to the trivial permutation.

Another argument pointing at the subtlety of defining deformed loop amplitudes
originates in the complicated branch-cut structure of the integrands. The naive
generalization of the Hankel contour leads to ill-defined integrals, as
discussed in \secref{sec:bubbles}

Without deformation, however, the $\rop$-operator formalism leads to exactly
the integrands written in ref.~\cite{ArkaniHamed:2012nw}, which is demonstrated
in \secref{sec:oneloopcalc}. 

Furthermore, we would like to point out that the correspondence between Yangian
invariants and permutations will break down at loop level. Indeed, for four
points there are only $4! = 24$ permutations and clearly at a loop level high
enough, this will be smaller than the amount of terms comprising the amplitude.

I'd change to: Finally, we would like to remark that the spectral parameters in
the R-operators do not seem to provide a straightforward regularization of loop
integrals; not even if we break Yangian invariance only mildly by keeping the
arguments of the R-operators general. This is due to the fact that the
$\rop$-operators have no mass dimension and consequently do not provide an
immediate tool to regulate the IR behavior of loop amplitudes. This argument
could be circumvented by considering an appropriate contour of integration (as
remarked in~\cite{StaudacherAmpl}).

Nevertheless, it would be very useful to further investigate the issue of loop
amplitudes in an algebraic language: modifications of the monodromy matrix could
still pave the way towards an understanding of the regularization of loop
amplitudes.

\paragraph{Acknowledgments.}

We would like to thank Niklas Beisert, James Drummond, Jan Plefka, and Cristian
Vergu for useful discussions and especially Tomasz \L{}ukowski and Matthias
Staudacher for discussions about the contours of integration.  The work of MdL
and MR is partially supported by grant no.\ 200021-137616 from the Swiss
National Science Foundation.

\appendix
\section{Eigenvalue property}\label{app:EigenV}

In this section we prove that eqns.~\eqref{eq:Rop1L4P} and \eqref{eq:Rop1L4Pb}
are eigenstates of the monodromy matrix with eigenvalues given in
\eqref{eq:EigenVLoop4}. In \cite{Chicherin:2013ora,Broedel:2014pia} it was
shown that eigenstates of the monodromy matrix respect dihedral symmetry. We
will use this and prove that the aforementioned states are eigenstates of
$\mathrm{T}$ by showing that they are eigenstates of the shifted monodromy
matrix
\begin{align}
\mathrm{T}_{s} = \lop_2(u_2)\lop_3(u_3)\lop_4(u_4)\lop_1(u_1).
\end{align}
We will discuss the procedure for eqn.~\eqref{eq:Rop1L4P} in detail; the
computation for eqn.~\eqref{eq:Rop1L4Pb} is completely analogous. First, we use the
rule that two $\rop$-operators commute for appropriate indices
\begin{align}\label{eq:Rpermute}
  &\rop_{ab}(u)\rop_{cd}(v) = \rop_{cd}(v)\rop_{ab}(u), &&~ \mathrm{if}~ a\neq d
  ~\mathrm{and}~ b \neq c,
\end{align}
to rewrite eqn.~\eqref{eq:Rop1L4P} as
\begin{align}
\mathcal{A}_{4;2}^{(1)}=
\rop_{41}(u_{14})\rop_{23}(u_{32})\rop_{43}(u_{43})\rop_{23}(u_{23})
\rop_{34}(u_{24})\rop_{21}(u_{21})\rop_{12}(u_{24})\rop_{23}(u_{14})
\Omega_{++--}.
\end{align}
It is quickly checked from eqns.~\eqref{eq:RLYBE} and \eqref{eq:RLYBEinv} that
$\mathrm{T}_{s}$ can be commuted through the $\rop$-operators up to the last three.
There we encounter the problem that indices one and two are not adjacent for
the shifted monodromy matrix. However, we can use one of the so-called
RR$\delta$-rules from \cite{Broedel:2014pia}
\begin{align}
  &\rop_{ab}(u)\rop_{bc}(v)\dl{a}\dl{b}\dlb{c}=
  \rop_{bc}(v-u)\rop_{ca}(-u)\dlb{a}\dl{b}\dl{c},
\end{align}
together with eqn.~\eqref{eq:Rpermute} to find the following way of expressing
eqn.~\eqref{eq:Rop1L4P}
\begin{align}
\mathcal{A}_{4;2}^{(1)}=
\rop_{41}(u_{14})\rop_{23}(u_{32})\rop_{43}(u_{43})\rop_{23}(u_{23})
\rop_{34}(u_{24})\rop_{13}(u_{14})\rop_{23}(u_{12})\rop_{23}(u_{21})
\Omega_{++--}.
\end{align}
Since to the right of $\rop_{13}$ there is no $\rop$-operator with index four, we
find that all operators have neighboring indices and consequently this is an
eigenstate of the monodromy matrix. In particular we find
\begin{align}
\mathrm{T}_s\, \mathcal{A}_{4;2}^{(1)} = (u_1+\half)(u_2-\half)(u_3+\half)(u_4-\half)\mathcal{A}_{4;2}^{(1)}.
\end{align}
The exact same considerations work for eqn.~\eqref{eq:Rop1L4Pb} as well since they
only differ by a parity-flip of the first $\rop$-operator, which does not spoil the
property that it is an eigenstate of shifted monodromy matrix. This results in 
\begin{align}
\mathrm{T}_s\, \mathcal{B}_{4;2}^{(1)} = (u_1-\half)(u_2-\half)(u_3+\half)(u_4+\half)\mathcal{B}_{4;2}^{(1)}.
\end{align}
We see that the eigenvalues are simply related by interchanging
$u_1\leftrightarrow u_4$. Because the central charges are vanishing for both
states, we find that the eigenvalues of $\mathrm{T}_s$ and the normal monodromy
matrix $\mathrm{T}$ coincide. This proves eqn.~\eqref{eq:EigenVLoop4}.

\begin{bibtex}[\jobname]

@article{Broedel:2014pia,
  author         = "Broedel, Johannes and de Leeuw, Marius and Rosso, Matteo",
  title          = "{A dictionary between R-operators, on-shell graphs and
                    Yangian algebras}",
  year           = "2014",
  eprint         = "1403.3670",
  archivePrefix  = "arXiv",
  primaryClass   = "hep-th",
  SLACcitation   = "
}

@article{Kanning:2014maa,
  author         = "Kanning, Nils and Lukowski, Tomasz and Staudacher,
                    Matthias",
  title          = "{A shortcut to general tree-level scattering amplitudes
                    in N=4 SYM via integrability}",
  year           = "2014",
  eprint         = "1403.3382",
  archivePrefix  = "arXiv",
  primaryClass   = "hep-th",
  SLACcitation   = "
}

@article{ArkaniHamed:2009dg,
  author        = {Arkani-Hamed, Nima and Bourjaily, Jacob and Cachazo, Freddy and Trnka, Jaroslav},
  title         = {{Unification of residues and Grassmannian dualities}},
  journal       = {JHEP},
  year          = {2011},
  volume        = {1101},
  pages         = {049},
  archiveprefix = {arXiv},
  doi           = {10.1007/JHEP01(2011)049},
  eprint        = {0912.4912},
  primaryclass  = {hep-th},
  slaccitation  = {
}

@article{ArkaniHamed:2010kv,
  author        = {Arkani-Hamed, Nima and Bourjaily, Jacob L. and Cachazo, Freddy and Caron-Huot, Simon and Trnka, Jaroslav},
  title         = {{The all-loop integrand for scattering amplitudes in planar N=4 SYM}},
  journal       = {JHEP},
  year          = {2011},
  volume        = {1101},
  pages         = {041},
  archiveprefix = {arXiv},
  doi           = {10.1007/JHEP01(2011)041},
  eprint        = {1008.2958},
  primaryclass  = {hep-th},
  slaccitation  = {
}

@article{ArkaniHamed:2012nw,
  author        = {Arkani-Hamed, Nima and Bourjaily, Jacob L. and Cachazo, Freddy and Goncharov, Alexander B. and Postnikov, Alexander and others},
  title         = {{Scattering amplitudes and the positive Grassmannian}},
  year          = {2012},
  archiveprefix = {arXiv},
  eprint        = {1212.5605},
  primaryclass  = {hep-th},
  slaccitation  = {
}

@article{ArkaniHamed:2009vw,
  author        = {Arkani-Hamed, Nima and Cachazo, Freddy and Cheung, Clifford},
  title         = {{The Grassmannian origin of dual superconformal invariance}},
  journal       = {JHEP},
  year          = {2010},
  volume        = {1003},
  pages         = {036},
  archiveprefix = {arXiv},
  doi           = {10.1007/JHEP03(2010)036},
  eprint        = {0909.0483},
  primaryclass  = {hep-th},
  slaccitation  = {
}

@article{ArkaniHamed:2009dn,
  author        = {Arkani-Hamed, Nima and Cachazo, Freddy and Cheung, Clifford and Kaplan, Jared},
  title         = {{A duality for the S matrix}},
  journal       = {JHEP},
  year          = {2010},
  volume        = {1003},
  pages         = {020},
  archiveprefix = {arXiv},
  doi           = {10.1007/JHEP03(2010)020},
  eprint        = {0907.5418},
  primaryclass  = {hep-th},
  slaccitation  = {
}

@article{ArkaniHamed:2008gz,
  author        = {Arkani-Hamed, Nima and Cachazo, Freddy and Kaplan, Jared},
  title         = {{What is the simplest Quantum Field Theory?}},
  journal       = {JHEP},
  year          = {2010},
  volume        = {1009},
  pages         = {016},
  archiveprefix = {arXiv},
  doi           = {10.1007/JHEP09(2010)016},
  eprint        = {0808.1446},
  primaryclass  = {hep-th},
  slaccitation  = {
}

@article{Beisert:2014qba,
  author        = {Beisert, Niklas and Broedel, Johannes and Rosso, Matteo},
  title         = {{On Yangian-invariant regularisation of deformed on-shell diagrams in N=4 super-Yang--Mills theory}},
  year          = {2014},
  archiveprefix = {arXiv},
  eprint        = {1401.7274},
  primaryclass  = {hep-th},
  slaccitation  = {
}

@article{Brandhuber:2008pf,
  author        = {Brandhuber, Andreas and Heslop, Paul and Travaglini, Gabriele},
  title         = {{A note on dual superconformal symmetry of the N=4 super Yang--Mills S-matrix}},
  journal       = {Phys.Rev.},
  year          = {2008},
  volume        = {D78},
  pages         = {125005},
  archiveprefix = {arXiv},
  doi           = {10.1103/PhysRevD.78.125005},
  eprint        = {0807.4097},
  primaryclass  = {hep-th},
  reportnumber  = {QMUL-PH-08-15},
  slaccitation  = {
}

@article{Brink:1982pd,
  author        = {Brink, Lars and Lindgren, Olof and Nilsson, Bengt E.W.},
  title         = {{N=4 Yang--Mills theory on the light cone}},
  journal       = {Nucl.Phys.},
  year          = {1983},
  volume        = {B212},
  pages         = {401},
  doi           = {10.1016/0550-3213(83)90678-8},
  reportnumber  = {GOTEBORG-82-21},
  slaccitation  = {
}

@article{Brink:1982wv,
  author        = {Brink, Lars and Lindgren, Olof and Nilsson, Bengt E.W.},
  title         = {{The ultraviolet finiteness of the N=4 Yang--Mills theory}},
  journal       = {Phys.Lett.},
  year          = {1983},
  volume        = {B123},
  pages         = {323},
  doi           = {10.1016/0370-2693(83)91210-8},
  reportnumber  = {UTTG-1-82},
  slaccitation  = {
}

@article{Brink:1976bc,
  author        = {Brink, Lars and Schwarz, John H. and Scherk, Joel},
  title         = {{Supersymmetric Yang--Mills theories}},
  journal       = {Nucl.Phys.},
  year          = {1977},
  volume        = {B121},
  pages         = {77},
  doi           = {10.1016/0550-3213(77)90328-5},
  reportnumber  = {CALT-68-574},
  slaccitation  = {
}

@article{Britto:2004ap,
  author        = {Britto, Ruth and Cachazo, Freddy and Feng, Bo},
  title         = {{New recursion relations for tree amplitudes of gluons}},
  journal       = {Nucl.Phys.},
  year          = {2005},
  volume        = {B715},
  pages         = {499-522},
  archiveprefix = {arXiv},
  doi           = {10.1016/j.nuclphysb.2005.02.030},
  eprint        = {hep-th/0412308},
  primaryclass  = {hep-th},
  slaccitation  = {
}

@article{Britto:2005fq,
  author        = {Britto, Ruth and Cachazo, Freddy and Feng, Bo and Witten, Edward},
  title         = {{Direct proof of tree-level recursion relation in Yang--Mills theory}},
  journal       = {Phys.Rev.Lett.},
  year          = {2005},
  volume        = {94},
  pages         = {181602},
  archiveprefix = {arXiv},
  doi           = {10.1103/PhysRevLett.94.181602},
  eprint        = {hep-th/0501052},
  primaryclass  = {hep-th},
  slaccitation  = {
}

@article{Chicherin:2013ora,
  author         = {Chicherin, D. and Derkachov, S. and Kirschner, R.},
  title          = {{Yang--Baxter operators and scattering amplitudes in
                       N=4 super-Yang--Mills theory}},
  year           = {2013},
  eprint         = {1309.5748},
  archivePrefix  = {arXiv},
  primaryClass   = {hep-th},
  SLACcitation   = {
}

@article{Chicherin:2013sqa,
  author        = {Chicherin, D. and Kirschner, R.},
  title         = {{Yangian symmetric correlators}},
  journal       = {Nucl.Phys.},
  year          = {2013},
  volume        = {B877},
  pages         = {484-505},
  archiveprefix = {arXiv},
  doi           = {10.1016/j.nuclphysb.2013.10.006},
  eprint        = {1306.0711},
  owner         = {mrosso},
  primaryclass  = {math-ph},
  slaccitation  = {
  timestamp     = {2014.01.24},
  url           = {http://arxiv.org/abs/arXiv:1306.0711}
}

@article{Nandan:2012rk,
  author        = {Nandan, Dhritiman and Wen, Congkao},
  title         = {{Generating all tree amplitudes in N=4 SYM by inverse soft limit}},
  journal       = {JHEP},
  year          = {2012},
  volume        = {1208},
  pages         = {040},
  archiveprefix = {arXiv},
  doi           = {10.1007/JHEP08(2012)040},
  eprint        = {1204.4841},
  primaryclass  = {hep-th},
  reportnumber  = {QMUL-PH-12-09},
  slaccitation  = {
}

@article{Dixon:2010ik,
  author         = {Dixon, Lance J. and Henn, Johannes M. and Plefka, Jan and
                    Schuster, Theodor},
  title          = {{All tree-level amplitudes in massless QCD}},
  journal        = {JHEP},
  volume         = {1101},
  pages          = {035},
  doi            = {10.1007/JHEP01(2011)035},
  year           = {2011},
  eprint         = {1010.3991},
  archivePrefix  = {arXiv},
  primaryClass   = {hep-ph},
  reportNumber   = {CERN-PH-TH-2010-230, SLAC-PUB-14278, HU-EP-10-57},
  SLACcitation   = {
}

@article{Drummond:2008vq,
  author        = {Drummond, J.M. and Henn, J. and Korchemsky, G.P. and Sokatchev, E.},
  title         = {{Dual superconformal symmetry of scattering amplitudes in N=4 super-Yang--Mills theory}},
  journal       = {Nucl.Phys.},
  year          = {2010},
  volume        = {B828},
  pages         = {317-374},
  archiveprefix = {arXiv},
  doi           = {10.1016/j.nuclphysb.2009.11.022},
  eprint        = {0807.1095},
  primaryclass  = {hep-th},
  reportnumber  = {LAPTH-1257-08, LPT-ORSAY-08-60},
  slaccitation  = {
}

@article{Drummond:2009fd,
  author        = {Drummond, James M. and Henn, Johannes M. and Plefka, Jan},
  title         = {{Yangian symmetry of scattering amplitudes in N=4 super Yang--Mills theory}},
  journal       = {JHEP},
  year          = {2009},
  volume        = {0905},
  pages         = {046},
  archiveprefix = {arXiv},
  doi           = {10.1088/1126-6708/2009/05/046},
  eprint        = {0902.2987},
  primaryclass  = {hep-th},
  reportnumber  = {HU-EP-09-06, LAPTH-1308-09},
  slaccitation  = {
}

@article{Ferro:2012xw,
  author        = {Ferro, Livia and Lukowski, Tomasz and Meneghelli, Carlo and Plefka, Jan and Staudacher, Matthias},
  title         = {{Harmonic R-matrices for scattering amplitudes and spectral regularization}},
  journal       = {Phys.Rev.Lett.},
  year          = {2013},
  volume        = {110},
  pages         = {121602},
  number        = {12},
  archiveprefix = {arXiv},
  doi           = {10.1103/PhysRevLett.110.121602},
  eprint        = {1212.0850},
  primaryclass  = {hep-th},
  reportnumber  = {HU-EP-12-50, HU-MATHEMATIK:14-2012, DESY-12-228, ZMP-HH-12-26, AEI-2012-198, -AEI-2012-198},
  slaccitation  = {
}

@article{Ferro:2013dga,
  author        = {Ferro, Livia and Lukowski, Tomasz and Meneghelli, Carlo and Plefka, Jan and Staudacher, Matthias},
  title         = {{Spectral parameters for scattering amplitudes in N=4 super Yang--Mills theory}},
  year          = {2013},
  archiveprefix = {arXiv},
  eprint        = {1308.3494},
  primaryclass  = {hep-th},
  reportnumber  = {HU-MATHEMATIK-2013-12, HU-EP-13-33, AEI-2013-235, DESY-13-488, --ZMP-HH-13-15},
  slaccitation  = {
}

@article{Frassek:2013xza,
  author         = {Frassek, Rouven and Kanning, Nils and Ko, Yumi and Staudacher, Matthias},
  title          = {{Bethe ansatz for Yangian invariants: towards super-Yang--Mills scattering amplitudes}},
  year           = {2013},
  eprint         = {1312.1693},
  archivePrefix  = {arXiv},
  primaryClass   = {math-ph},
  reportNumber   = {HU-MATHEMATIK-2013-14, HU-EP-13-34, AEI-2013-234, DCPT-13-47},
  SLACcitation   = "
}

@article{Gliozzi:1976qd,
  author        = {Gliozzi, F. and Scherk, Joel and Olive, David I.},
  title         = {{Supersymmetry, supergravity theories and the dual spinor model}},
  journal       = {Nucl.Phys.},
  year          = {1977},
  volume        = {B122},
  pages         = {253-290},
  doi           = {10.1016/0550-3213(77)90206-1},
  reportnumber  = {CERN-TH-2253},
  slaccitation  = {
}

@article{Howe:1983sr,
  author        = {Howe, Paul S. and Stelle, K.S. and Townsend, P.K.},
  title         = {{Miraculous ultraviolet cancellations in supersymmetry made manifest}},
  journal       = {Nucl.Phys.},
  year          = {1984},
  volume        = {B236},
  pages         = {125},
  doi           = {10.1016/0550-3213(84)90528-5},
  reportnumber  = {ICTP-82-83-20},
  slaccitation  = {
}

@article{Mandelstam1982,
  author        = {Mandelstam, S.},
  title         = {{Light cone superspace and the finiteness of the N=4 model}},
  year          = {1982},
  reportnumber  = {CERN-TH-3385},
  slaccitation  = {
}

@article{Nair:1988bq,
  author        = {Nair, V.P.},
  title         = {{A current algebra for some gauge theory amplitudes}},
  journal       = {Phys.Lett.},
  year          = {1988},
  volume        = {B214},
  pages         = {215},
  doi           = {10.1016/0370-2693(88)91471-2},
  reportnumber  = {CU-TP-408},
  slaccitation  = {
}

@article{Witten:2003nn,
  author        = {Witten, Edward},
  title         = {{Perturbative gauge theory as a string theory in twistor space}},
  journal       = {Commun.Math.Phys.},
  year          = {2004},
  volume        = {252},
  pages         = {189-258},
  archiveprefix = {arXiv},
  doi           = {10.1007/s00220-004-1187-3},
  eprint        = {hep-th/0312171},
  primaryclass  = {hep-th},
  slaccitation  = {
}

@article{Postnikov:2006kva,
  author        = {Postnikov, Alexander},
  title         = {{Total positivity, Grassmannians, and networks}},
  year          = {2006},
  archiveprefix = {arXiv},
  eprint        = {math/0609764},
  primaryclass  = {math.CO},
  slaccitation  = {
}

@article{Drinfeld:1985rx,
  author         = "Drinfeld, V. G.",
  title          = "Hopf algebras and the quantum Yang--Baxter equation",
  journal        = "Sov. Math. Dokl.",
  volume         = "32",
  year           = "1985",
  pages          = "254-258",
  SLACcitation   = "
}

@article{Drinfeld:1986in,
  author         = "Drinfeld, V. G.",
  title          = "Quantum groups",
  journal        = "J. Math. Sci.",
  volume         = "41",
  year           = "1988",
  pages          = "898",
  doi            = "10.1007/BF01247086",
  SLACcitation   = "
}

@article{Takhtajan:1979iv,
  author        = "Takhtajan, L. A. and Faddeev, L. D.",
  title         = "{The quantum method of the inverse problem and the
                    Heisenberg XYZ model}",
  journal       = "Russ.Math.Surveys",
  volume        = "34",
  pages         = "11-68",
  year          = "1979",
  SLACcitation  = "
}

@article{Kulish:1980ii,
  author        = "Kulish, P. P. and Sklyanin, E. K.",
  title         = "{On the solution of the Yang--Baxter equation}",
  journal       = "J.Sov.Math.",
  volume        = "19",
  pages         = "1596-1620",
  doi           = "10.1007/BF01091463",
  year          = "1982",
  SLACcitation  = "
}
@article{Faddeev:1987ih,
  author        = "Faddeev, L. D. and Reshetikhin, N. {\relax Yu}. and Takhtajan,
                    L.A.",
  title         = "{Quantization of Lie groups and Lie algebras}",
  journal       = "Leningrad Math.J.",
  volume        = "1",
  pages         = "193-225",
  year          = "1990",
  reportNumber  = "LOMI-E-14-87",
  SLACcitation  = "
}

@article{Khoroshkin:1996fy,
  author        = "Khoroshkin, S. and Lebedev, D. and Pakuliak, S.",
  title         = "{Intertwining operators for the central extension of the
                    Yangian double}",
  year          = "1996",
  eprint        = "q-alg/9602030",
  archivePrefix = "arXiv",
  primaryClass  = "q-alg",
  reportNumber  = "DFTUZ-95-28, ITEP-TH-15-95",
  SLACcitation  = "
}

@article{Khoroshkin:1996fz,
  author        = "Khoroshkin, S.M.",
  title         = "{Central extension of the Yangian double}",
  year          = "1996",
  eprint        = "q-alg/9602031",
  archivePrefix = "arXiv",
  primaryClass  = "q-alg",
  SLACcitation  = "
}

@article{Bern:2004ky,
  author         = "Bern, Zvi and Del Duca, Vittorio and Dixon, Lance J. and
                    Kosower, David A.",
  title          = "{All non-maximally-helicity-violating one-loop
                    seven-gluon amplitudes in N=4 super-Yang--Mills theory}",
  journal        = "Phys.Rev.",
  volume         = "D71",
  pages          = "045006",
  doi            = "10.1103/PhysRevD.71.045006",
  year           = "2005",
  eprint         = "hep-th/0410224",
  archivePrefix  = "arXiv",
  primaryClass   = "hep-th",
  reportNumber   = "SLAC-PUB-10810, UCLA-04-TEP-43, DFTT-26-2004,
                    DCPT-04-136, IPPP-04-68, SACLAY-SPHT-T04-131,
                    NSF-KITP-04-114",
  SLACcitation   = "
}

@article{MacKay:2004tc,
  author         = "MacKay, N.J.",
  title          = "{Introduction to Yangian symmetry in integrable field
                    theory}",
  journal        = "Int.J.Mod.Phys.",
  volume         = "A20",
  pages          = "7189-7218",
  doi            = "10.1142/S0217751X05022317",
  year           = "2005",
  eprint         = "hep-th/0409183",
  archivePrefix  = "arXiv",
  primaryClass   = "hep-th",
  reportNumber   = "ESI-1514",
  SLACcitation   = "
}

@article{Khoroshkin:1994uk,
  author         = "Khoroshkin, S.M. and Tolstoi, V.N.",
  title          = "{Yangian double and rational R matrix}",
  year           = "1994",
  eprint         = "hep-th/9406194",
  archivePrefix  = "arXiv",
  primaryClass   = "hep-th",
  SLACcitation   = "
}

@article{Beisert:2014hya,
  author         = "Beisert, Niklas and de Leeuw, Marius",
  title          = "{The RTT-Realization for the deformed $\mathfrak{gl}(2|2)$ Yangian}",
  year           = "2014",
  eprint         = "1401.7691",
  archivePrefix  = "arXiv",
  primaryClass   = "math-ph",
  SLACcitation   = "
}

@article{Beisert:2011pn,
  author         = "Beisert, Niklas and Schwab, Burkhard U.W.",
  title          = "{Bonus Yangian symmetry for the planar S-Matrix of N=4
                    super Yang--Mills}",
  journal        = "Phys.Rev.Lett.",
  volume         = "106",
  pages          = "231602",
  doi            = "10.1103/PhysRevLett.106.231602",
  year           = "2011",
  eprint         = "1103.0646",
  archivePrefix  = "arXiv",
  primaryClass   = "hep-th",
  reportNumber   = "AEI-2011-006",
  SLACcitation   = "
}

@article{Mason:2009qx,
  author         = "Mason, L.J. and Skinner, David",
  title          = "{Dual superconformal invariance, momentum twistors and
                    Grassmannians}",
  journal        = "JHEP",
  volume         = "0911",
  pages          = "045",
  doi            = "10.1088/1126-6708/2009/11/045",
  year           = "2009",
  eprint         = "0909.0250",
  archivePrefix  = "arXiv",
  primaryClass   = "hep-th",
  SLACcitation   = "
}

@article{Bullimore2010,
  author         = "Bullimore, Mathew and Mason, L.J. and Skinner, David",
  title          = "{MHV diagrams in momentum twistor space}",
  journal        = "JHEP",
  volume         = "1012",
  pages          = "032",
  doi            = "10.1007/JHEP12(2010)032",
  year           = "2010",
  eprint         = "1009.1854",
  archivePrefix  = "arXiv",
  primaryClass   = "hep-th",
  SLACcitation   = "
}

@article{Green1982,
  author       = {Green, Michael B. and Schwarz, John H. and Brink, Lars},
  title        = {{N=4 Yang--Mills and N=8 supergravity as limits of string theories}},
  journal      = {Nucl.Phys.},
  year         = {1982},
  volume       = {B198},
  pages        = {474-492},
  doi          = {10.1016/0550-3213(82)90336-4},
  reportnumber = {CALT-68-880},
  slaccitation = {
}

@article{Bern1994,
  author        = {Bern, Zvi and Dixon, Lance J. and Dunbar, David C. and Kosower, David
	A.},
  title         = {{One loop $n$-point gauge theory amplitudes, unitarity and collinear
	limits}},
  journal       = {Nucl.Phys.},
  year          = {1994},
  volume        = {B425},
  pages         = {217-260},
  archiveprefix = {arXiv},
  doi           = {10.1016/0550-3213(94)90179-1},
  eprint        = {hep-ph/9403226},
  primaryclass  = {hep-ph},
  reportnumber  = {SLAC-PUB-6415, SACLAY-SPH-T-94-20, UCLA-TEP-94-4, SWAT-94-17},
  slaccitation  = {
  timestamp     = {2014.04.01}
}

@article{Kosower:2010yk,
  author         = "Kosower, D.A. and Roiban, R. and Vergu, C.",
  title          = "{The six-Point NMHV amplitude in maximally supersymmetric
                    Yang--Mills Theory}",
  journal        = "Phys.Rev.",
  volume         = "D83",
  pages          = "065018",
  doi            = "10.1103/PhysRevD.83.065018",
  year           = "2011",
  eprint         = "1009.1376",
  archivePrefix  = "arXiv",
  primaryClass   = "hep-th",
  reportNumber   = "SACLAY-IPHTT10-096, BROWN-HET-1607",
  SLACcitation   = "
}

@article{StaudacherAmpl,
  author         = "{Staudacher, M., talk given at the \textrm{XIX} Itzykson Conference, \textit{Amplitudes 2014, a Claude Itzykson memorial conference}, June 10 - 13, 2014, CEA Saclay}",
  title          = "",
  url = "http://ipht.cea.fr/en/Meetings/Itzykson2014/talks/Staudacher-Itzykson19.pdf",
}

\end{bibtex}

\bibliographystyle{nb}
\bibliography{\jobname}

\end{document}